\documentclass[prx, twocolumn,longbibliography]{revtex4-1}

\usepackage{bbm}
\usepackage{amsmath}
\usepackage{hyperref}
\usepackage{graphicx}
\usepackage{graphics}
\usepackage{amssymb}
\usepackage{mathtools}
\usepackage{physics}
\usepackage{csquotes}
\usepackage{graphicx}
\usepackage{color}
\usepackage{comment}
\usepackage[dvipsnames]{xcolor}

\graphicspath{{Figs/}}

\newcommand{\pp}{{{}^{++}}}
\newcommand{\p}{{{}^{+}}}
\newcommand{\mm}{{{}^{--}}}
\newcommand{\m}{{{}^{-}}}

\begin{document}

\title{Driven dynamics of a quantum dot electron spin coupled to bath of higher-spin nuclei}

\author{Arian Vezvaee$^1$} 
\email{avezva@vt.edu}
\author{Girish Sharma$^2$} 
\author{Sophia E. Economou$^1$}
\author{Edwin Barnes$^1$}
\email{efbarnes@vt.edu}
\affiliation{$^1$Department of Physics, Virginia Tech, Blacksburg, Virginia 24061, USA\\$^2$School of Basic Sciences, Indian Institute of Technology Mandi, India}

%%%%%%%%%%%%%%%%%%%%%%%%%
%%%%%%%%%%%%%%%%%%%%%%%%%

\begin{abstract}
The interplay of optical driving and hyperfine interaction between an electron confined in a quantum dot and its surrounding nuclear spin environment produces a range of interesting physics such as mode-locking. In this work, we go beyond the ubiquitous spin 1/2 approximation for nuclear spins and present a comprehensive theoretical framework for an optically driven electron spin in a self-assembled quantum dot coupled to a nuclear spin bath of arbitrary spin. Using a dynamical mean-field approach, we compute the nuclear spin polarization distribution with and without the quadrupolar coupling. We find that while hyperfine interactions drive dynamic nuclear polarization and mode-locking, quadrupolar couplings counteract these effects. The tension between these mechanisms is imprinted on the steady-state electron spin evolution, providing a way to measure the importance of quadrupolar interactions in a quantum dot. Our results show that higher-spin effects such as quadrupolar interactions can have a significant impact on the generation of dynamic nuclear polarization and how it influences the electron spin evolution.
\end{abstract}

%%%%%%%%%%%%%%%%%%%%%%%%%
%%%%%%%%%%%%%%%%%%%%%%%%%

\maketitle

%%%%%%%%%%%%%%%%%%%%%%%%%%%%%%%%%%%%%%%%%%%%%%%%%%
%%%%%%%%%%%%%%%%%%%%%%%%%%%%%%%%%%%%%%%%%%%%%%%%%%

\section{Introduction}\label{sec-intro}

Spins in self-assembled quantum dots (QDs) are under intense investigation for a variety of quantum information applications, including quantum information processing, quantum communication, and quantum transduction~\cite{Ladd_2010,Liu,Lauk,Wehnereaam9288}. The relatively long coherence times, fast controllability~\cite{Economou2006,Greilich_2009,Makhonin2011}, and good photon emission properties of these systems~\cite{Ding2016,Buckley2012,Wei2014,Senellart2017} make them promising candidates for achieving high-quality spin-photon interfaces and for producing large-scale multi-photon entangled states~\cite{Lindner,Buterakos,Russo,hilaire}. The deterministic generation of these multi-photon entangled states has been demonstrated experimentally using the dark excitonic states of QDs~\cite{Schwartz}.

While optically controlled quantum dot spins offer a wide range of technological possibilities, hyperfine (HF) interactions between the confined spin and its surrounding nuclear spin bath have been a major impediment. This interaction is the main source of decoherence in these systems, and it also causes spectral wandering and inhomogeneities in quantum dot ensembles, aspects that have been researched extensively over the past two decades~\cite{Urbaszek,Huang2010,Huang2010,Merkulov2002,Yao2006,Stockill,Coish,Cywinski2009PRL,Cywinski2009,Barnes2012,Khaetskii,Yao2006,Bortz,Gammon2001,Chekhovich2020,Gammon96,Gammon97,Chekhovich,Wuest,Chekhovich2010,Ladd,Vink,Schering2019,Schering2020,Beugeling,Kleinjohann,Welander2014,Chekhovich2015,Kloeffel,Maletinsky,Chekhovich2012nat,Krebs2010,Chekhovich2013nat,Chekhovich2010PRB,Greilich2007PRB,Munsch,Prechtel,Schering2020arxiv}. However, many works have shown that the state of the bath, and consequently its deleterious effects, can be influenced by driving the electron spin. For example, several experiments have shown that driving can generate dynamic nuclear polarization (DNP), an effect that has been observed in self-assembled QDs~\cite{Greilich2006,Greilich,Carter,Varwig2012,Varwig2013,Varwig2014,Xu,Bracker,Latta_2009,Hogele,Latta2011,Dominguez} and also in other systems such as gated QDs~\cite{Foletti_2009,Reilly,Bluhm,Nichol}, quantum wires~\cite{Pribiag2013} and in bulk materials~\cite{Tiane,Rej}, findings that have been supported by a number of theory works~\cite{BarnesPRL2011,Yang2013,EconomouPRB2014,Stano,Nutz,Sharma2017,Sharma2019,Neder,Neder2014,Rudner,Danon2009,Oulton,Bulutay}. An important example of DNP in self-assembled QDs is the mode-locking experiment of Ref.~\cite{Greilich}, where an ensemble of QD electron spins becomes synchronized with a periodic train of optical pulses as a consequence of DNP. Continuous-wave laser
driving of the electron has been shown to create DNP in QDs as well, leading to interesting phenomena such as the line-dragging effect, i.e. the locking of an optical QD transition to the frequency of the laser~\cite{Xu,Hogele,Latta_2009,Yang_2012,Nutz}. Owing to the long coherence times of nuclear spins, DNP has been proposed for applications such as quantum memories~\cite{Taylor,Denning}, which has recently been demonstrated experimentally~\cite{Gangloff62}.   

Although most of the fully quantum mechanical theoretical treatments of the hyperfine decoherence problem allow for nuclei with spin greater than 1/2~\cite{Cywinski2009,Cywinski2009PRL,Coish}, studies of the driven, hyperfine-induced generation of DNP have mostly focused on spin 1/2 nuclei to reduce the computational complexity of the problem \cite{BarnesPRL2011,EconomouPRB2014,Rudner,Danon,Greilich,Danon2009}. The latter works typically rely on either stochastic equations or rate equations to solve for the nuclear polarization distribution. While solving the feedback problem for spin 1/2 nuclear baths can yield qualitative insights about DNP experiments, the quantitative accuracy of such models is limited by the fact that the most commonly studied semiconductor QDs are in materials such as InAs or GaAs, which contain nuclei of spin $I > 1/2$. In addition to artificially reducing the size of the bath Hilbert space, assuming spin 1/2 nuclei also ignores effects such as quadrupolar interactions, which are only present for $I>1/2$. There do exist a few theoretical works that allow for $I>1/2$~\cite{Huang2010,Yang2013,Beugeling}. Specifically, Huang and Hu~\cite{Huang2010} studied DNP arising from hyperfine interactions with the spin 3/2 arsenic nuclei in InGaAs by making use of Fermi's golden rule; however, only qualitative agreement with experiment was achieved due to the need to introduce phenomenological parameters. Yang and Sham~\cite{Yang2013} presented a general framework for nuclei of arbitrary total spin by unifying the stochastic and rate-equation approaches. In this work they focused on a drift feedback loop (which allows for a possible bias in nuclear spin-flip processes) and obtained a Fokker-Planck equation for the polarization of the bath. Although this framework captures line-dragging and other DNP phenomena seen in experiments, it has only been established for continuous-wave driving, and so it is not immediately applicable to experiments with periodic driving such as the mode-locking experiment of Ref.~\cite{Greilich}. Theoretical works that have specifically focused on mode-locking type experiments have either assumed $I=1/2$ nuclear baths~\cite{BarnesPRL2011,EconomouPRB2014} or utilized semiclassical methods \cite{Schering2019,Schering2020}. While such approaches have been successful in reproducing qualitative features seen in experiments including dynamic nuclear polarization and mode-locking, it remains an outstanding challenge to develop a more quantitatively accurate description of the driven electron-nuclear spin system. Allowing for higher spin is also important for capturing additional qualitative behavior that can arise from quadrupolar interactions.

In this paper, we develop a quantum, non-perturbative framework to solve the dynamics of an optically driven electron spin coupled to a bath of $I>1/2$ nuclear spins. We focus on DNP feedback mechanisms that arise from driving the electron with a periodic train of optical pulses while it is subject to hyperfine interactions with a nuclear spin bath, as in the mode-locking experiment \cite{Greilich,Greilich2006}. Here, we also consider the effect of quadrupolar interactions. To compute DNP and its effect on the evolution of the electron spin, we use an approach based on dynamical maps and kinetic equations introduced in Refs.~\cite{BarnesPRL2011,EconomouPRB2014}, but, importantly, we generalize the formalism to higher nuclear spin and treat the problem non-perturbatively. Our framework provides a self-consistent description of the feedback loop between the driven electron and DNP. 

We compute the nuclear spin polarization distribution and its influence on the electron spin evolution for spin 1 and spin 3/2 baths and compare the results to the $I=1/2$ case. Our approach is able to treat bath sizes of up to thousands of nuclear spins in the $I=1/2$ and $I=1$ cases, and up to several hundred spins in the $I=3/2$ case. Although evidence of mode-locking is seen in all three cases, we find that quadrupolar interactions act to suppress mode-locking for $I>1/2$, especially when the angle between the principal strain axis and the applied magnetic field is large. We also find that while HF interactions can produce a significant bath polarization that grows linearly with the number of nuclei for $I>1/2$, quadrupolar interactions work to counteract this buildup of DNP. We further show that the relative importance of quadrupolar effects grows as the magnitude of the applied magnetic field is increased. The competition between HF and quadrupolar interactions imprints clear signatures in the steady-state electron spin evolution, providing an experimental tool to measure the strength of quadrupolar couplings in a QD. Our results show that accounting for higher nuclear spin is important not only for quantitative accuracy, but also for capturing important qualitative features of the DNP process in driven QD systems.

The paper is structured as follows. In Sec.~\ref{sec-overview}, we present an overview of our framework and briefly describe each step of the calculation. In Sec.~\ref{sec-formalism}, we lay out the approach in detail for arbitrary nuclear spin $I$ and construct the equations that govern DNP for $I=1/2$, 1, and 3/2 nuclear spin baths. We present an analytical solution for the steady-state nuclear spin polarization distribution for $I=1/2$. In Sec.~\ref{sec:polarization_distr_feedback}, we numerically compute steady-state polarization distributions for $I=1$ and 3/2 and compare the results to the $I=1/2$ solution for various parameter choices. We also study the effect of DNP on the electron spin evolution. We conclude in Sec.~\ref{sec-conclusions}.

\begin{figure}
\includegraphics[scale=.6]{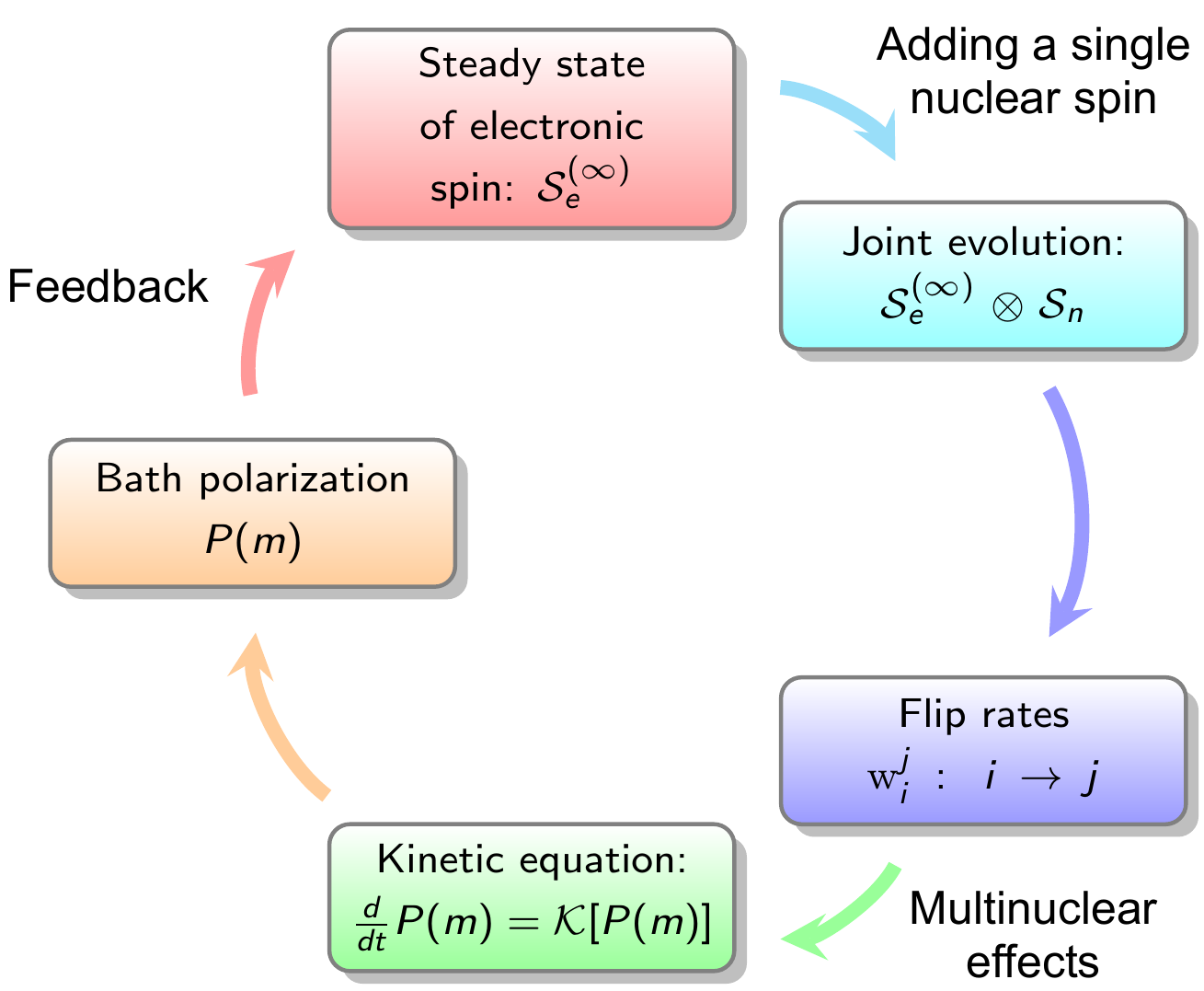}
\caption{~Schematic depiction of the self-consistent formalism we use to model DNP with feedback. We exploit a hierarchy of timescales to first solve for the joint evolution of the electron coupled to a single nuclear spin. Under a Markovian approximation, the electron spin state is reset after each drive period. The resulting nuclear spin evolution yields nuclear spin-flip rates that are then fed into a kinetic equation governing the dynamics of the multi-nuclear spin polarization distribution. The flip rates depend on the effective electron spin precession frequency, including the Overhauser field contribution for self-consistency. The solution to the kinetic equation is then used to update the electron steady state, closing the feedback loop.}
\label{fig-framework}
\end{figure}

%%%%%%%%%%%%%%%%%%%%%%%%%%%%%%%%%%%%%%%%%%%%%%%%%%
%%%%%%%%%%%%%%%%%%%%%%%%%%%%%%%%%%%%%%%%%%%%%%%%%%

\section{Overview of the theoretical  framework}\label{sec-overview}

Before we describe our approach in detail, we first give an overview of the general strategy and main ingredients. This will hopefully better orient the reader for what follows and also highlight the generality of the approach, which could potentially be adapted to other time-dependent many-body problems. We still, however, frame this overview in the context of the electron-nuclear problem in QDs for the sake of concreteness and to make it readily apparent how the discussion here fits with the rest of the paper. Our framework is summarized in Fig.~\ref{fig-framework}. It closely follows the approach introduced in Refs.~\cite{BarnesPRL2011,EconomouPRB2014}, but we outline each step in detail to keep the discussion self-contained and to make it clearer which parts must be modified to allow for higher nuclear spin.

Our focus in this work is on developing a theory that describes QD experiments in which a single electron is periodically pumped by a train of optical pulses~\cite{Greilich,Carter,Carter2011}. Each pulse excites the electron to a trion state (a bound state of an electron and an exciton), which then decays back to the electronic ground state manifold via spontaneous emission.
The underlying physical mechanism behind the formation of DNP can be understood as follows. Imagine that the electron spin starts in a pure (polarized) state and the nuclear spins are in a totally mixed (unpolarized) state. The HF interaction then transfers angular momentum from the electron onto the nuclei, creating DNP. In the absence of driving, this would lead to only a modest nuclear spin polarization, and this polarization would be short-lived because it would eventually be transferred back to the electron via the HF interaction. However, the laser pulses periodically reset the electron spin to a polarized state, enabling a net transfer of angular momentum from the laser, through the electron, and onto the nuclei. It is this transfer process that we aim to describe with our framework.

We are dealing with a system that is both open and driven. An efficient way to treat non-unitary evolution is to use dynamical maps~\citep{nielsen,BarnesPRL2011,EconomouPRB2014,Jordan1961,Jordan2004}. In this approach, the non-unitary evolution of a system from an initial state $\rho$ to a final state $\rho'$ is implemented by applying a set of operators and summing the results:
\begin{equation}
\rho^\prime = \sum_k E_k \rho E^\dag _k.\label{eq:kraus}
\end{equation}
The operators $E_k$ are known as Kraus operators, and they constitute a generalization of the usual unitary operators that evolve closed quantum systems to the case of non-unitary evolution in open systems. The condition $\sum_k E^\dag _k E_k   = \mathbbm{1}$ ensures that the trace of the density matrix is always unity. The advantage of Kraus operators is that they allow one to incorporate effects due to the transient occupation of excited states using operators that live purely in the ground space of the system. In the present problem, we use these operators to describe the effect of each optical pulse on the electron spin state. The entire process of optical excitation, subsequent decay, and rotation is captured by an appropriate set of Kraus operators (given in the next section) without having to explicitly include excited states or a photonic bath into the formalism. The dynamical map description works well so long as the population returns regularly to the electron spin ground states, as is the case for the periodic driving used in the mode-locking experiments.

Given a set of Kraus operators that describe the evolution of the driven electron, the next step in our theoretical framework is to find the steady-state of the electron spin. Of course, we are interested in the case where the electron spin is coupled to a nuclear spin bath through HF interactions (which are described in detail in the next section) while it is being driven. Under the condition that the electron is being pumped fast enough (which indeed is the case for the mode-locking experiments \cite{Greilich}), the electron reaches its steady state on a much faster timescale compared to the electron-nuclear interaction dynamics and the electron spin decoherence time. This allows us to use a Markovian approximation in which we first solve for the driven electron steady state and then incorporate the effects due to the electron-nuclear couplings on top of this.

To bring the nuclei into the framework, we first solve for the joint evolution of one nuclear spin coupled to the driven electron spin. Although the HF interaction generates unitary dynamics, this is disrupted periodically by the pulses, and this in turn leads to an effective non-unitary dynamical map for the nuclear spin that depends on the electron steady state under the Markovian approximation. We extract nuclear spin-flip rates from this effective nuclear spin evolution operator; these rates provide information about the movement of population between the different nuclear spin levels.

We calculate the steady state of the entire nuclear spin bath using a rate equation that depends on the spin-flip rates obtained from the single-nucleus solution. A critical step is that we build in self-consistent system-environment feedback by modifying the flip rates. To understand this, we first need to describe the Overhauser effect~\cite{Overhauser}, which is the main feedback mechanism between the electron and nuclei. A polarized nuclear spin bath acts as an effective magnetic field and therefore shifts the Zeeman frequency of the electron. However, the interaction between the electron and the nuclear spin bath is reciprocal; not only will the state of the electron change under the Overhauser field, but the nuclear spins will also be affected by the Knight field~\cite{Knight}, i.e., the effective magnetic field due to polarization of the electron. The Knight field is given by the electron steady state spin vector, and so it enters into the nuclear spin flip rates, as explained above. The electron steady state (and hence the Knight field) in turn depends on the total magnetic field, which includes the Overhauser field due to nuclear polarization. These interdependencies constitute a complete feedback loop that must be treated self-consistently. We do this by making the nuclear spin-flip rates depend on the net nuclear polarization of the bath. The steady-state of the rate equation then gives the polarization distribution of the nuclear spin bath with feedback included. Finally, we use this nuclear polarization distribution to perform the Overhauser shift on the Zeeman frequency of the electron and update the nuclear-bath-averaged electron spin steady-state self-consistently.

The framework we have just outlined can be thought of as a self-consistent dynamical mean-field approach.
In the following section, we describe each step of our formalism as it applies to the periodically driven electron-nuclear problem in full detail. Our method is quite general and can be applied to baths of any nuclear spin. We focus on the cases $I=1/2$, 1, and 3/2 to illustrate the various steps.

%%%%%%%%%%%%%%%%%%%%%%%%%%%%%%%%%%%%%%%%%%%%%%%%%%
%%%%%%%%%%%%%%%%%%%%%%%%%%%%%%%%%%%%%%%%%%%%%%%%%%

\section{Self-consistent dynamical mean-field formalism and results}\label{sec-formalism}

%\subsection{Hamiltonian}\label{sec:ham}

\begin{figure}
\includegraphics[scale=1]{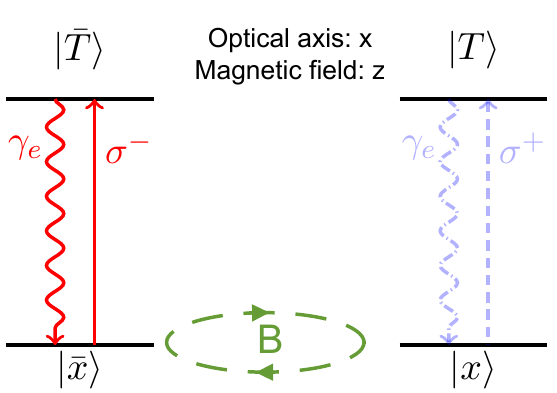}
\caption{The relevant level structure in the mode-locking experiments. $\ket{x}$ and $\ket{\bar{x}}$ are the electron spin states along the optical axis. These states are coupled by an external magnetic field along the $z$ direction. Circularly polarized light excites the ground electron spin states to excited trion levels $\ket{T}$ and $\ket{\bar{T}}$ with angular momentum projections $+3/2$ and $-3/2$, respectively. The selection rules are such that each ground state couples to only one excited state. The trion states decay via spontaneous emission with rate $\gamma_e$. In this work, we focus on left-circularly polarized driving.}
\label{fig-setup}
\end{figure}

The full Hamiltonian of the nuclear spin bath and the driven electron is given by
\begin{equation} \label{eq-full-H}
H(t) = H_{0,e} + H_{0,n} + H_c(t) + H_{res} + H_{HF} + H_Q.
\end{equation}
Here, $H_{0,e}$ describes the electronic degrees of freedom in the QD in the absence of driving:
\begin{equation}
H_{e,0} = \omega_e \hat{S}_z + \omega_{\bar{T}}
|\bar{T}\rangle\langle \bar{T}|,
\end{equation}
where $\omega_e$ is the electron spin Zeeman frequency, $\hat{S}_z$ is the spin operator in the electronic ground space, and $\omega_{\bar{T}}$ is the energy of the trion state $|\bar{T}\rangle$. We take the magnetic field to be oriented along the $z$ direction, while the optical axis lies in the $x$ direction (see Fig.~\ref{fig-setup}). We neglect the second trion level $\ket{T}$ in $H_{0,e}$ because it is not excited by the laser polarization we are considering. This driving is described by the Hamiltonian
\begin{equation}
    H_c(t)=\Omega(t) |\bar{x}\rangle\langle \bar{T}| + h.c.,
\end{equation}
where we assume the drive laser is left-circularly polarized (red arrow in Fig.~\ref{fig-setup}) with periodic temporal profile $\Omega(t+T_R)=\Omega(t)$, so that each pulse couples the electron spin state $\ket{\bar{x}}$ to the trion state $\ket{\bar T}$. The latter decays via spontaneous emission with rate $\gamma_e$. This process arises from interactions with a photonic bath, which is represented by the term $H_{res}$. We do not give an explicit expression for this term as it is not explicitly considered in what follows. The Zeeman splitting of the nuclear spins is given by $H_{0,n}=\omega_n \sum_i \hat I_z^i$.

The HF interaction is given by the contact term:
\begin{equation} \label{eq-HF}
H_{HF} = \sum _{i=1}^N A_i \hat{S}_z \hat{I}^i_z + \sum _{i=1}^N{A_i}/{2}(\hat{S}_+\hat{I}^i_-+\hat{S}_-\hat{I}^i_+),
\end{equation}
where $N$ is the number of nuclei that interact appreciably with the electron. The first term is referred to as the Overhauser term, and it gives rise to an effective magnetic field seen by the electron spin in the case of nonzero nuclear spin polarization. The second term generates flip-flop interactions under which the electron spin flips with a nuclear spin. These terms are responsible for transferring angular momentum from the electron onto the nuclei, while the Overhauser term is the primary mechanism for feedback between the nuclear spin polarization and the electron spin evolution. The HF couplings $A_i$ are determined by the magnitude of the electronic wave function at the location of the nuclear spin $I^i$. However, on timescales short compared to $N/\mathcal{A}\sim\mu$s, where $\mathcal{A}$ is the total HF interaction energy, the variations in these couplings do not significantly affect the electron spin evolution \cite{BarnesPRLBox}. Here, we focus on fast optical driving where the electron reaches a steady state over a timescale of about 100 ns \cite{EconomouPRB2014}, which allows us to make the \enquote{box model} approximation in which all the HF couplings are taken equal: $A_i= A\equiv\mathcal{A}/N$~\cite{Barnes2011,BarnesPRLBox}.

The quadrupolar interaction is given by~\citep{abragam,slichter} 
\begin{equation}\label{eq-HQ}
H_Q = \sum_{i=1}^N\frac{\nu_Q^i}{2}\left(\hat{I}_{z'}^i{}^2-\frac{I(I+1)}{3}\right).
\end{equation}
This interaction occurs due to the coupling of the nuclear quadrupole moment to electric field gradients caused by strain in the semiconductor lattice, and it is only present for $I>1/2$. The presence of quadrupolar interactions has lead to striking phenomena in various types of experiments conducted in QDs. A few examples include the anomalous Hanle effect~\cite{Krebs2010} and suppression of spin diffusion~\citep{Maletinsky}. Line-dragging phenomena have also been associated with the presence of quadrupolar interactions~\citep{Hogele,Latta_2009,Yang2013}. The coupling strength $\nu_Q$ is referred to as the nuclear quadrupole resonance frequency, which is estimated to be around 2.8 MHz for As~\cite{Maletinsky}. The quadrupole resonance frequency generally depends on the local strain in the vicinity of each nuclear spin, and so it generally varies across the material. Here, we assume that the strain remains roughly constant over the QD, and so we take all the frequencies to be equal: $\nu_Q^i=\nu_Q$. The operator $\hat{I}_{z'}$ in Eq.~(\ref{eq-HQ}) is the component of the nuclear spin operator along the principal axis of the electric field gradient. Our focus will be on the case of QDs with cylindrical symmetry in which the electric field gradient makes an angle $\theta$ with the magnetic field. Therefore, we have $\hat{I}_{z'}=\hat{I}_z\cos\theta+\hat{I}_x\sin\theta$, which then gives~\cite{abragam}:
\begin{eqnarray}
H_Q &=& \frac{\nu_Q}{2}\sum_{i=1}^N \Big[(\hat{I}^i_{z})^2 \cos^2\theta -\frac{I(I+1)}{3} \nonumber \\
&+& (\hat{I}^i_{z}\hat{I}^i_{x}+\hat{I}^i_{x}\hat{I}^i_{z})\sin \theta \cos \theta + (\hat{I}^i_{x})^2 \sin^2\theta\Big].
\end{eqnarray}
When $\theta=0$, $H_Q$ creates non-uniform energy spacings between the nuclear spin levels. For $\theta\ne0$, $H_Q$ has the additional effect of driving $\Delta m_I=\pm1$ and $\Delta m_I=\pm2$ nuclear spin-flip transitions, where $m_I$ is the eigenvalue of $\hat I_z$. Notice that the rate for $\Delta m_I=\pm1$ transitions is maximal at $\theta=\pi/4$, while the rate for $\Delta m_I=\pm2$ transitions is largest for $\theta=\pi/2$, which is also the value of $\theta$ where the non-uniformity in the energy level spacings is zero. Thus, we see that the role of $H_Q$ changes as $\theta$ varies from 0 to $\pi/4$, and from $\pi/4$ to $\pi/2$. Because $H_Q$ is $\pi$-periodic in $\theta$, it suffices to focus on the range $0\le\theta\le\pi/2$.

In the following subsections, we carry out the various steps of the formalism as outlined above in Sec.~\ref{sec-overview}. In Sec.~\ref{sec-Kr-form}, we introduce the Kraus operators that describe the optical pumping process and subsequent spontaneous emission generated by each pulse. In Sec.~\ref{sec-eSS}, we then combine these with the Larmor precession between pulses to construct a dynamical map that evolves the electron spin vector over one complete driving cycle (with duration $T_R$). This dynamical map is used to obtain the steady state of the electron spin vector. In Sec.~\ref{sec-joint}, we derive a dynamical map that describes the effective evolution of a single nuclear spin hyperfine-coupled to the electron and subject to quadrupolar effects, where we use the Markovian approximation to freeze the electron in its steady state as described above. We use this dynamical map to compute the nuclear spin-flip rates in Sec.~\ref{sec-fliprates}. We obtain non-perturbative analytical expressions for these flip rates for an $I=1/2$ bath and for $I=1$ and $I=3/2$ baths in the case of no quadrupolar interactions. Results with the quadrupolar interaction included are obtained numerically. We take into account multinuclear effects in Sec.~\ref{sec-ke} by constructing kinetic equations for $I=1/2$, 1, and 3/2 that govern the time evolution of the nuclear spin polarization distribution of the entire bath. In these equations, we include Overhauser shifts in the flip rates to incorporate dynamical feedback effects between the electron and nuclei. The steady states of these equations describe the DNP that is created through the combination of optical pumping of the electron and hyperfine flip-flops. These equations are then solved in various cases in Sec.~\ref{sec:polarization_distr_feedback}, where we also describe how the resulting polarization distributions in turn modify the evolution of the electron spin.

%%%%%%%%%%%%%%%%%%%%%%%%%%%%%%%%%%

\subsection{Kraus operators for optical pumping of the electron} \label{sec-Kr-form}

The existence of a hierarchy of timescales in mode-locking experiments allows us to first solve for the electron spin dynamics without having to include nuclear spin effects. This is due to the fact that the nuclear spin dynamics are slow compared to those of the electron. Given that the nuclear spins are the main source of decoherence for the electron, this means we can also neglect electron spin decoherence effects. In addition, the optical pumping and spontaneous emission are fast compared to the pulse period, $\gamma_e T_R \gg 1$, which ensures that the excited population returns fully to the ground state before the next pulse comes. This allows us to treat the evolution of the electron over each period in terms of a dynamical map that acts only on the electron spin ground state subspace, as in Eq.~\eqref{eq:kraus}.

The Kraus operators, $E_k$, that make up the dynamical map can be found by explicitly computing the non-unitary part of the evolution for an arbitrary initial density matrix and comparing the initial and final density matrices. To compute the non-unitary part of the evolution due to the sequence of pulses $H_c(t)$, we only need the electronic parts of the full Hamiltonian in Eq.~(\ref{eq-full-H}): $H_e(t) = H_{0,e}+H_c(t)$. The fact that the pulse is much shorter than the spin precession period allows us to ignore the precession during the action of the pulse. Therefore $\ket{\bar{x}}$ and $\ket{\bar T}$ can be considered as an effective two-level system, where the evolution operator due to the pulse in the $\ket{x}$, $\ket{\bar{x}}$, $\ket{\bar T}$ basis is
\begin{equation}
    U_p=\left[
\begin{array}{ccc}
1 & 0  &  0  \\
0  & u_{\bar{x}\bar{x}} & -u_{\bar{T}\bar{x}}^* \\
0 & u_{\bar{T}\bar{x}} & u_{\bar{x}\bar{x}}^*
\end{array}
\right].
\end{equation}
After the pulse, a fraction $|u_{\bar{T}\bar{x}}|^2$ of the population remains in the trion state. We can describe the decay of this population due to spontaneous emission using the Liouville-von Neumann equation with appropriately chosen Lindblad operators $\mathcal{L}$: $\dot \rho=i[\rho,H_{0,e}] + \mathcal{L}(\rho)$, where the first term includes the Larmor precession of the ground spin states during the decay.
Solving this equation for an arbitrary initial state then yields the following Kraus operators in the $\ket{x}$, $\ket{\bar{x}}$ basis~\cite{EconomouPRB2014}:
\begin{equation} \label{eq-kraus-e}
   E_1 = \begin{bmatrix} 1 & 0  \\0 & q\end{bmatrix}, \quad  E_2 = \begin{bmatrix}0 & \;\;\;a_1  \\0 & -a_2\end{bmatrix},\quad  E_3 = \begin{bmatrix} 0 & 0  \\
0 & \kappa\end{bmatrix},
\end{equation}
where $q = u_{\bar{x}\bar{x}}\equiv q_o e^{i\phi}$, $a_1 =\omega_e \sqrt{(1-q_o^2)/2(4\gamma_e^2+\omega_e^2)}$, $a_2 = i \gamma_e\sqrt{2}\sqrt{(1-q_o^2)/(4\gamma_e^2+\omega_e^2)}$, and $\kappa = \sqrt{1-q_o^2-a_1^2-|a_2|^2}$. These Kraus operators guarantee the unity of the trace of the density matrix by satisfying $\sum_k E^\dag _k E_k  = \mathbbm{1}$. The parameter $q_o$ quantifies the amount of population remaining in the spin state $\ket{\bar{x}}$ after the pulse is applied, and $\phi$ is the angle about the $x$ axis by which the pulse rotates the electron spin. These two parameters can be computed given a specific pulse shape, but in the following we leave these parameters arbitrary.

%%%%%%%%%%%%%%%%%%%%%%%%%%%%%%%%%%

\subsection{Electron spin steady state} \label{sec-eSS}

We can use the Kraus operators from above to compute the electron spin steady state. Rather than work directly with the Kraus operators, it is more convenient to switch to the spin vector (SV) representation, especially since finding the steady state requires applying the Kraus operators an infinite number of times. In general, a SV $S$ transforms under non-unitary evolution as follows:
\begin{equation}\label{eq-SV-evolution}
S'=YS+K,
\end{equation}
where $Y$ is a matrix that generally both rotates and shrinks the SV, while $K$ corresponds to the non-unital part of the evolution (i.e., a loss or gain of population in the subspace described by $S$). If $K$ is nonzero, then a nontrivial steady state is possible. As shown in Ref.~\cite{EconomouPRB2014}, for spin 1/2 these quantities can be obtained from the Kraus operators using the following formulas:
\begin{eqnarray}\label{eq-k-y}
K_i &=& \frac{1}{2} \text{Tr} \sum_k \hat\sigma_i \mathcal{E}_k  \mathcal{E}_k^\dagger, \\
 Y_{ij} &=& \frac{1}{2}\text{Tr} \sum_k \hat\sigma_i \mathcal{E}_k \hat\sigma_j \mathcal{E}_k^\dagger,
 \\
 \nonumber
\end{eqnarray}
where the $\hat\sigma_i$ are Pauli matrices. In the case of the mode-locking experiment, the Kraus operators $\mathcal{E}_k$ evolve the electron spin over one period, that is, they include both the non-unitary dynamics ($E_k$) generated by a pulse and also the unitary precession under the magnetic field over time $T_R$: $\mathcal{E}_k=E_ke^{-i\omega_eT_R\hat{S}_z}$. To find the steady state, it is convenient to combine both $Y$ and $K$ into a single  $4\times 4$ matrix:
\begin{equation}
\mathcal{Y}_e= \begin{bmatrix}
1 & 0 & 0 & 0 \\
K_{x} & Y_{xx} & Y_{xy} & Y_{xz} \\
K_{y} & Y_{yx} & Y_{yy} & Y_{yz} \\
K_{z} & Y_{zx} & Y_{zy} & Y_{zz}
\end{bmatrix},\label{eq:4x4Y}
\end{equation}
where the evolution of the electron SV over one period is now given by $\mathcal{S}'_e =\mathcal{Y}_e \mathcal{S}_e $. Here, the first component of the 4-component SV $\mathcal{S}_e$ is always fixed to 1, while the remaining three components constitute the usual spin 1/2 SV. In this representation it is easy to see that the steady state $\mathcal{S}^{ss}_{e}=(1,S^{ss}_{e,x},S^{ss}_{e,y},S^{ss}_{e,z})$ is the eigenvector of $\mathbbm{1}-\mathcal{Y}_e$ with eigenvalue zero. Transforming the Kraus operators of Eq.~(\ref{eq-kraus-e}) from the $x$ basis to the $z$ basis, plugging the result into Eq.~\eqref{eq:4x4Y}, and computing the null vector of $\mathcal{Y}_e$ leads to the following steady state electron SV~\cite{EconomouPRB2014}:
\begin{widetext}
\begin{eqnarray}
S_{e,x}^{ss}&{=}&\frac{a_1 \left(a_1 q_o \left(q_o-\cos\phi\right) \cos \left(\omega _e T_R\right)-i a_2 \left(q_o \cos\phi-1\right) \sin \left(\omega _e
   T_R\right)-a_1 q_o\cos\phi+a_1\right)}{\left(a_1^2+q_o^2-1\right) \cos \left(\omega _e T_R\right)-a_1 q_o \cos \phi  \left[i a_2
   \sin \left(\omega _e T_R\right)+a_1 \cos \left(\omega _e T_R\right)+a_1\right]+i a_1 a_2 \sin \left(\omega _e T_R\right)+\left(a_1^2-1\right)
   q_o^2+1},\nonumber\\
S_{e,y}^{ss}&{=}&\frac{a_1 \left(a_1 q_o \left(\cos\phi-q_o\right) \sin \left(\omega _e T_R\right)-i a_2 \left(q_o \cos\phi-1\right) \left(\cos
   \left(\omega _e T_R\right)-1\right)\right)}{\left(a_1^2+q_o^2-1\right) \cos \left(\omega _e T_R\right)-a_1 q_o \cos\phi \left[i a_2 \sin
   \left(\omega _e T_R\right)+a_1 \cos \left(\omega _e T_R\right)+a_1\right]+i a_1 a_2 \sin \left(\omega _e T_R\right)+\left(a_1^2-1\right)
   q_o^2+1},\nonumber\\
S_{e,z}^{ss}&{=}&\frac{a_1 q_o \sin\phi \left(a_1 \sin \left(\omega _e T_R\right)-i a_2 \left(\cos \left(\omega _e
   T_R\right)-1\right)\right)}{\left(a_1^2+q_o^2-1\right) \cos \left(\omega _e T_R\right)-a_1 q_o \cos\phi \left[i a_2 \sin \left(\omega _e
   T_R\right)+a_1 \cos \left(\omega _e T_R\right)+a_1\right]+i a_1 a_2 \sin \left(\omega _e T_R\right)+\left(a_1^2-1\right) q_o^2+1}.\nonumber\\&&\label{eq-ss-e-SV}
\end{eqnarray}
\end{widetext}
These are the components of the electron SV immediately after each pulse. The steady state at other times during the driving period can be obtained by rotating this vector about the $z$ axis by angle $\omega_eT_R$ (to account for the Larmor precession).

%%%%%%%%%%%%%%%%%%%%%%%%%%%%%%%%%%

\subsection{Effective dynamical map for one nuclear spin} \label{sec-joint}

Now that we have the electron spin steady state (Eq.~\eqref{eq-ss-e-SV}), we can proceed to construct an effective dynamical map for a single nuclear spin. We do this by first constructing the evolution operator in the SV representation that describes the joint evolution of the electron and nuclear spins over one driving period. We then apply the Markovian approximation and reset the electron spin to its steady state at the end of the period. Tracing out the electron then leaves an effective dynamical map for the nuclear spin.

To start, we define the nuclear SV using a basis of Hermitian matrices $\hat\lambda_{k}$ of dimension $2I+1$, where $k=1,\ldots,(2I+1)^2$. We choose the first $2I+1$ of these matrices to be diagonal, each with a single nonzero component equal to one. The remaining $2I(2I+1)$ matrices each have two nonzero components, and these matrices are purely real or purely imaginary. For example, in the case of $I=3/2$, we have 16 basis matrices:
\begin{widetext}
\begin{eqnarray}
    &&\hat\lambda_{k,ab}=\delta_{ak}\delta_{bk}, \quad k=1\ldots4,\nonumber\\
   &&\hat\lambda_{5,ab}=\tfrac{1}{\sqrt{2}}(\delta_{a1}\delta_{b2}+\delta_{a2}\delta_{b1}),\quad \hat\lambda_{6,ab}=\tfrac{-i}{\sqrt{2}}(\delta_{a1}\delta_{b2}-\delta_{a2}\delta_{b1}),\quad \hat\lambda_{7,ab}=\tfrac{1}{\sqrt{2}}(\delta_{a1}\delta_{b3}+\delta_{a3}\delta_{b1}),\nonumber\\
   && \hat\lambda_{8,ab}=\tfrac{-i}{\sqrt{2}}(\delta_{a1}\delta_{b3}-\delta_{a3}\delta_{b1}),\quad\hat\lambda_{9,ab}=\tfrac{1}{\sqrt{2}}(\delta_{a1}\delta_{b4}+\delta_{a4}\delta_{b1}),\quad \hat\lambda_{10,ab}=\tfrac{-i}{\sqrt{2}}(\delta_{a1}\delta_{b4}-\delta_{a4}\delta_{b1}),\nonumber\\
   &&\hat\lambda_{11,ab}=\tfrac{1}{\sqrt{2}}(\delta_{a2}\delta_{b3}+\delta_{a3}\delta_{b2}),\quad \hat\lambda_{12,ab}=\tfrac{-i}{\sqrt{2}}(\delta_{a2}\delta_{b3}-\delta_{a3}\delta_{b2}),\quad\hat\lambda_{13,ab}=\tfrac{1}{\sqrt{2}}(\delta_{a2}\delta_{b4}+\delta_{a4}\delta_{b2}),\nonumber\\
   && \hat\lambda_{14,ab}=\tfrac{-i}{\sqrt{2}}(\delta_{a2}\delta_{b4}-\delta_{a4}\delta_{b2}),\quad\hat\lambda_{15,ab}=\tfrac{1}{\sqrt{2}}(\delta_{a3}\delta_{b4}+\delta_{a4}\delta_{b3}),\quad \hat\lambda_{16,ab}=\tfrac{-i}{\sqrt{2}}(\delta_{a3}\delta_{b4}-\delta_{a4}\delta_{b3}).
\end{eqnarray}
\end{widetext}
These matrices are normalized such that $\hbox{Tr}[\hat\lambda_j\hat\lambda_k]=\delta_{jk}$.
Denoting the nuclear spin density matrix as $\rho_n$, the components of the nuclear SV $\mathcal{S}_n$ are then given by
\begin{equation}
    \mathcal{S}_{n,k}=\hbox{Tr}[\rho_n\lambda_k].
\end{equation}
Note that the populations, $\rho_{n,ii}$, are the first four components of $\mathcal{S}_n$. We will see that this feature simplifies the process of computing flip rates.

Let us denote the density matrix that describes the total electron-nuclear spin state at the beginning of a driving period by $\varrho$. We expand this in terms of an operator basis formed from tensor products of the nuclear spin operators $\hat \lambda_{k}$ with the electron spin Pauli matrices $\hat \sigma_j$:
\begin{equation} \label{eq-def-G}
\hat G_{(2I+1)^2j+k}=\hat \sigma_j\otimes \hat{\lambda}_{k},
\end{equation}
with $j=0,..,3$, $k=1,...,(2I+1)^2$, and where we define $\hat \sigma_{0} = \mathbbm{1}_{2\times 2}$. We use this set of $4(2I+1)^2$ operators as a basis for the SV of the joint system: $\mathcal{S}_\ell=  \text{Tr}(\varrho \hat G_\ell)$. This SV evolves over one driving period according to $\mathcal{S}'=\mathcal{Y}\mathcal{S}$, where the SV evolution operator $\mathcal{Y}$ is given by
\begin{equation} \label{eq-def-Y-comp}
\mathcal{Y}_{\ell\ell'}=\frac{1}{2}\text{Tr}\left[\hat G_\ell \mathcal{U}  \hat G_{\ell'} \mathcal{U}^\dagger\right],
\end{equation}
where $\mathcal{U}=\exp\{-i(\omega_e\hat{S}_z+\omega_n\hat{I}_z+H_{HF}^{N=1}+H_Q^{N=1})T_R\}$ describes the joint evolution of the electron spin and single nuclear spin under precession and the HF and quadrupolar interactions.
At this point, we invoke the Markovian approximation: Because the electron reaches its steady state, $\mathcal{S}^{ss}_{e}$, quickly compared to the timescales for nuclear spin and HF dynamics, we reset the electron SV to its steady state value at the beginning/end of each period: $\mathcal{S}= \mathcal{S}^{ss}_{e}\otimes\mathcal{S}_n$. We then obtain an effective nuclear spin dynamical map, $\mathcal{Y}_n$, by acting with the full evolution operator, $\mathcal{Y}$, on the tensor product $\mathcal{S}^{ss}_{e}\otimes\mathcal{S}_n$ and reading off the coefficients of the components of the nuclear SV, $\mathcal{S}_n$, from the resulting $\mathcal{S}'$:
\begin{equation} 
\mathcal{Y}_{n,jk}=\frac{d}{d\mathcal{S}_{n,k}}\left[ \mathcal{Y} (\mathcal{S}^{ss}_{e}\otimes\mathcal{S}_n) \right]_j.\label{eq-def-Yn-comp}
\end{equation}
Here, $j,k=1,...,(2I+1)^2$, that is, we only retain the components of $\mathcal{S}'$ that correspond to the basis operators $\hat{G}_{k}=\mathbbm{1}_{2\times 2}\otimes\hat{\lambda}_{k}$, i.e., the components that correspond to purely nuclear spin degrees of freedom. Note that although the joint evolution operator $\mathcal{Y}$ describes unitary evolution, the nuclear spin dynamical map, $\mathcal{Y}_n$, implements non-unitary evolution. This non-unitarity is a consequence of the Markovian approximation, which is itself due to the non-unitary driving of the electron spin.

%%%%%%%%%%%%%%%%%%%%%%%%%%%%%%%%%%

\subsection{Single-nucleus flip rates}\label{sec-fliprates}

We can use the nuclear spin dynamical map, $\mathcal{Y}_n$, that we found in the previous subsection to find the flip rates for a single nuclear spin interacting with the electron spin. These flip rates govern the movement of population from one nuclear spin state to another. Such processes are described by the following kinetic equation:
\begin{equation}\label{eq-generalizedRateEqn}
\frac{d p_m}{dt} = \sum_{n \neq m} \mathrm{w}^m_n p_n - \sum_{n \neq m} \mathrm{w}^n_m p_m,
\end{equation}
where $p_m$ is the population of level $m$, and $\mathrm{w}^m_n$ is the rate to flip from state $n$ to $m$, which in general differs from the rate to flip from $m$ to $n$, $\mathrm{w}^n_m$. Which transitions are allowed depends on the type of interactions present in the Hamiltonian. For instance, the HF flip-flop terms only cause $\Delta m_I=\pm 1$ transitions, while the quadrupolar interaction also drives $\Delta m_I=\pm 2$ transitions. We can combine the rate equations \eqref{eq-generalizedRateEqn} into a matrix equation. We exemplify this in the $I=3/2$ case, where we denote the four states $\ket{+3/2}$, $\ket{+1/2}$, $\ket{-1/2}$, $\ket{-3/2}$, by the shorthand $\{++,+,-,--\}$. The matrix equation is then $\dot{\mathcal{P}} = \mathcal{M} \mathcal{P}$, where $\mathcal{P}=(p_{++},p_+,p_-,p_{--})$, and 
\begin{widetext}
\begin{equation} \label{eq-flip-rates-matrix}
\mathcal{M} =
 \begin{bmatrix} - (\mathrm{w}_\pp^\p + \mathrm{w}_\pp^\m +\mathrm{w}_{++}^{--})  & \mathrm{w}_\p^\pp & \mathrm{w}_\m^\pp & \mathrm{w}_{--}^{++} \\
  \mathrm{w}_\pp^\p & - (\mathrm{w}_\p^\pp + \mathrm{w}_\p^\m + \mathrm{w}_\p^\mm )   &  \mathrm{w}_\m^\p & \mathrm{w}_\mm^\p \\  
  \mathrm{w}_\pp^\m & \mathrm{w}_\p^\m& - (\mathrm{w}_\m^\pp + \mathrm{w}_\m^\p + \mathrm{w}_\m^\mm ) & \mathrm{w}_\mm^\m\\
  \mathrm{w}_{++}^{--} & \mathrm{w}_\p^\mm & \mathrm{w}_\m^\mm &  - (\mathrm{w}_\mm^\p + \mathrm{w}_\mm^\m +\mathrm{w}_{--}^{++})  \end{bmatrix}. 
\end{equation}
\end{widetext}
It is clear that this equation satisfies the condition that the sum of the components of the probability vector $\mathcal{P}$  should be unity at all times. This is guaranteed by the property that the sum of the rows of $\mathcal{M}$ vanishes.

\begin{figure*}
\includegraphics[scale=.65]{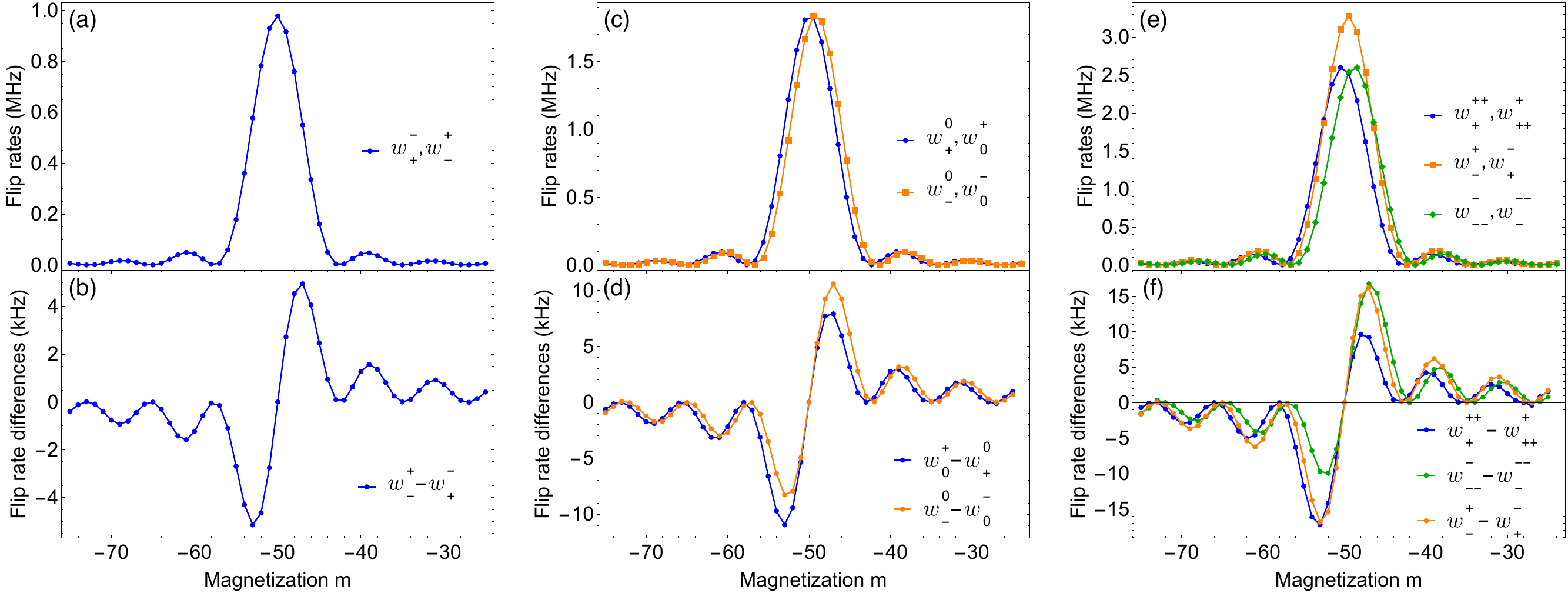}
\caption{Single-nucleus spin-flip rates for (a,b) $I=1/2$, (c,d) $I=1$, (e,f) $I=3/2$ as a function of the magnetization $m$ of the nuclear spin bath. Flip rates are shown in (a), (c), (e), while flip rate differences are shown in (b), (d), (f). The parameter values are
$T_R=13.2$~ns, $NA=10$~GHz, $N=1000$, $\omega_{e0}=0.5$~GHz, $\omega_n=-0.5$~MHz,  $\gamma_e=0.5$~GHz, $q_0=0.3$,
$\phi=-\pi/2$. For (c-f), we set the quadrupolar parameters to $\nu_Q = 2.8$~MHz and $\theta=0$. Only the nonzero flip rates are shown.}
\label{fig-fr}
\end{figure*}

To determine the flip rates, we need to connect the generic kinetic equation, Eq.~\eqref{eq-generalizedRateEqn}, to the nuclear spin evolution operator, Eq.~\eqref{eq-def-Yn-comp}, derived earlier. This can be done by starting from the evolution over one driving period:
\begin{equation} \label{eq-coarse-1}
\mathcal{S}_n(t+T_R) = \mathcal{Y}_n\mathcal{S}_n(t).
\end{equation}
The fact that the nuclear spin evolution is much slower than the driving period $T_R$ allows us to coarse-grain this equation to arrive at a continuous evolution equation:
\begin{equation}\label{eq-coarse-2}
\frac{d}{dt} \mathcal{S}_n = {1\over T_R}(\mathcal{Y}_n-\mathbbm{1})\mathcal{S}_n.
\end{equation}
Because we have defined $\mathcal{S}_n$ such that its first four components are just the populations of the nuclear spin states, we can identify this equation with $\dot{\cal{P}}=\cal{M}\cal{P}$ and therefore read off the flip-rate matrix components from the nuclear spin evolution matrix:
\begin{eqnarray}
\mathcal{M}_{ij}=\frac{1}{T_R}(\mathcal{Y}_n-\mathbbm{1})_{ij},\quad i,j=1...2I+1.
\end{eqnarray}
This allows us to read off the flip rates from the nuclear spin dynamical map. It is worth noting that $\mathcal{Y}_n$ contains not only terms that mix the populations of the different nuclear spin levels but also terms that mix populations and nuclear spin coherences. Here, we are neglecting the influence of the latter on the late-time populations. In numerical simulations, we find that these terms have a negligible effect on the flip rates. Moreover, they will be further suppressed by nuclear spin dephasing \cite{Makhonin2011,Chekhovich}, which happens quickly compared to nuclear spin flips.

In the case of $I=1/2$ nuclei, the flip rates can be obtained analytically following the above procedure:
\begin{equation}\label{eq:spin1/2fr}
\mathrm{w}_\pm=\frac{A^2(1\pm S_{e,z}^{ss})\sin^2(T_R\sqrt{(\omega_e-\omega_n)^2+A^2}/2)}{2T_R[(\omega_e-\omega_n)^2+A^2]},
\end{equation}
where we use the shorthand notation $\mathrm{w}_+\equiv \mathrm{w}_{-1/2}^{+1/2}$ and $\mathrm{w}_-\equiv \mathrm{w}_{+1/2}^{-1/2}$.
The flip rates for $I=1$ and $I=3/2$ can also be obtained analytically in the case of zero quadrupolar coupling, $\nu_Q=0$. In this case, there are four nonzero flip rates for $I=1$:
\begin{eqnarray}\label{eq:spin1fr}
\mathrm{w}_{-1}^0&=&\frac{A^2(1+S_{e,z}^{ss})\sin^2(T_R\Omega_-^{(1)}/2)}{T_R(\Omega^{(1)}_-)^2},\nonumber\\
\mathrm{w}_{0}^{-1}&=&\frac{A^2(1-S_{e,z}^{ss})\sin^2(T_R\Omega_-^{(1)}/2)}{T_R(\Omega^{(1)}_-)^2},\nonumber\\
\mathrm{w}_{0}^{+1}&=&\frac{A^2(1+S_{e,z}^{ss})\sin^2(T_R\Omega_+^{(1)}/2)}{T_R(\Omega^{(1)}_+)^2},\nonumber\\
\mathrm{w}_{+1}^{0}&=&\frac{A^2(1-S_{e,z}^{ss})\sin^2(T_R\Omega_+^{(1)}/2)}{T_R(\Omega^{(1)}_+)^2},
\end{eqnarray}
with
\begin{equation}
    \Omega_\pm^{(1)}=\sqrt{(\omega_e-\omega_n)^2\pm A(\omega_e-\omega_n)+9A^2/4},
\end{equation}
while there are six nonzero flip rates for $I=3/2$:
\begin{eqnarray}\label{eq:spin3/2fr}
\mathrm{w}_+^{++}&=&\frac{3A^2(1+S_{e,z}^{ss})\sin^2(T_R\Omega^{(3/2)}_{+1}/2)}{2T_R(\Omega^{(3/2)}_{+1})^2},\nonumber\\
\mathrm{w}_{++}^+&=&\frac{3A^2(1-S_{e,z}^{ss})\sin^2(T_R\Omega^{(3/2)}_{+1}/2)}{2T_R(\Omega^{(3/2)}_{+1})^2},\nonumber\\
\mathrm{w}_{-}^+&=&\frac{A^2(1+S_{e,z}^{ss})\sin^2(T_R\Omega^{(3/2)}_0/2)}{T_R(\Omega^{(3/2)}_0)^2},\nonumber\\
\mathrm{w}_{+}^-&=&\frac{A^2(1-S_{e,z}^{ss})\sin^2(T_R\Omega^{(3/2)}_0/2)}{T_R(\Omega^{(3/2)}_0)^2},\nonumber\\
\mathrm{w}_{--}^-&=&\frac{3A^2(1+S_{e,z}^{ss})\sin^2(T_R\Omega^{(3/2)}_{-1}/2)}{2T_R(\Omega^{(3/2)}_{-1})^2},\nonumber\\
\mathrm{w}_{-}^{--}&=&\frac{3A^2(1-S_{e,z}^{ss})\sin^2(T_R\Omega^{(3/2)}_{-1}/2)}{2T_R(\Omega^{(3/2)}_{-1})^2},
\end{eqnarray}
with
\begin{equation}
    \Omega^{(3/2)}_\eta=\sqrt{(\omega_e-\omega_n)^2+2\eta A(\omega_e-\omega_n)+4A^2}.
\end{equation}
In the absence of quadrupolar interactions, only $\Delta m_I=\pm1$ transitions (i.e., transitions between adjacent spin levels) are allowed, as follows directly from the form of the HF flip-flop interaction. When the quadrupolar coupling is nonzero, we can no longer obtain an analytical expression for the flip rates, but these are still easily obtained numerically by computing $\mathcal{Y}_n$ for specific parameter values.

\begin{figure*}
\includegraphics[scale=.65]{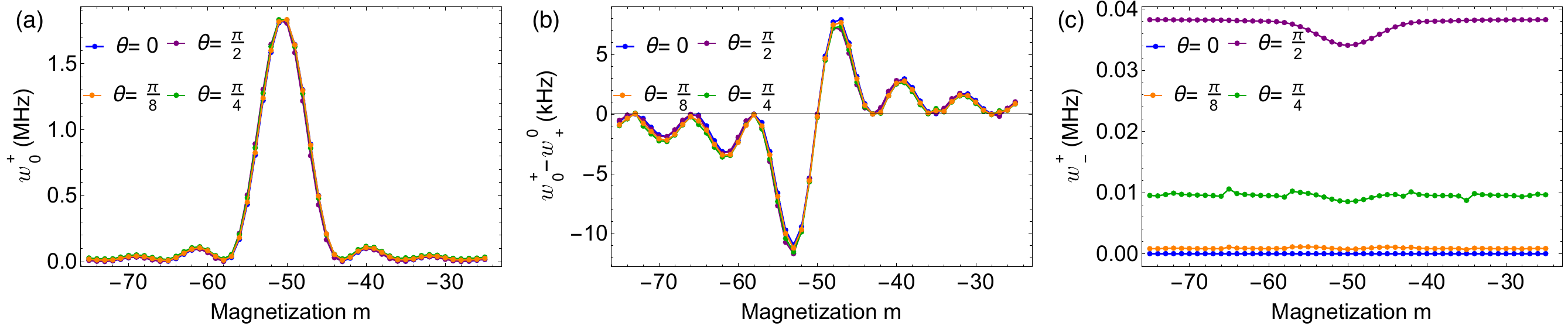}
\caption{Single-nucleus spin-flip rates as a function of nuclear spin bath magnetization $m$ for $I=1$ and for different values of the quadrupolar angle $\theta$. (a) Flip rate for the $\Delta m_I=1$ transition $\ket{0}\to\ket{+1}$. (b) Flip rate difference for the $\ket{0}\leftrightarrow\ket{+1}$ transitions. (c) Flip rate for the $\Delta m_I=2$ transition $\ket{-1}\to\ket{+1}$. The parameter values are
$T_R=13.2$~ns, $NA=10$~GHz, $N=1000$, $\omega_{e0}=0.5$~GHz, $\omega_n=-0.5$~MHz,  $\gamma_e=0.5$~GHz, $q_0=0.3$,
$\phi=-\pi/2$, $\nu_Q = 2.8$~MHz.}\label{fig-fr-one-nonzero-theta}
\end{figure*}

\begin{figure*}
\includegraphics[scale=.65]{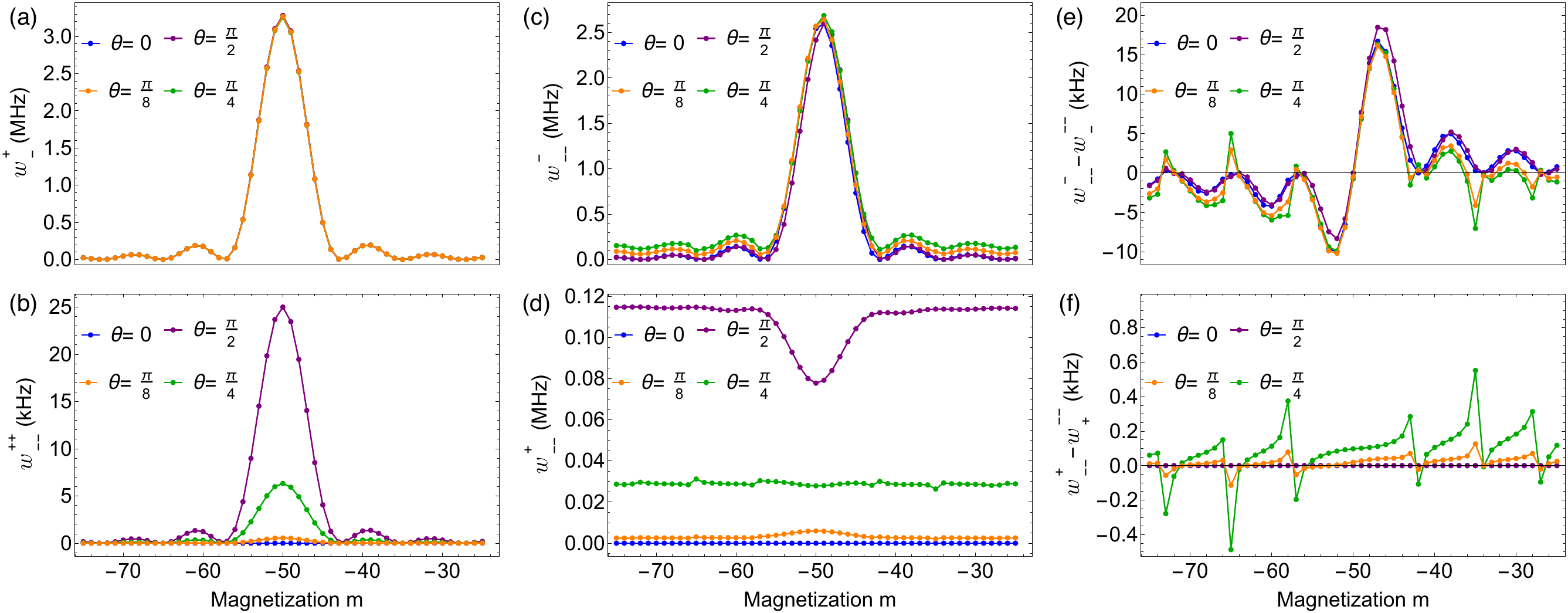}
\caption{Single-nucleus spin-flip rates as a function of nuclear spin bath magnetization $m$ for $I=3/2$ and for different values of the quadrupolar angle $\theta$. (a) Flip rate for the $\Delta m_I=1$ transition $\ket{-1/2}\to\ket{+1/2}$. (b) Flip rate for the $\Delta m_I=3$ transition $\ket{-3/2}\to\ket{+3/2}$. (c) Flip rate for the $\Delta m_I=1$ transition $\ket{-3/2}\to\ket{-1/2}$. (d) Flip rate for the $\Delta m_I=2$ transition $\ket{-3/2}\to\ket{+1/2}$. (e) Flip rate difference for the $\ket{-3/2}\leftrightarrow\ket{-1/2}$ transitions. (f) Flip rate difference for the $\ket{-3/2}\leftrightarrow\ket{+1/2}$ transitions. The parameter values are
$T_R=13.2$~ns, $NA=10$~GHz, $N=1000$, $\omega_{e0}=0.5$~GHz, $\omega_n=-0.5$~MHz,  $\gamma_e=0.5$~GHz, $q_0=0.3$,
$\phi=-\pi/2$, $\nu_Q = 2.8$~MHz.}\label{fig-fr-three-half-nonzero-theta}
\end{figure*}

Fig.~\ref{fig-fr} shows the dependence of the flip rates on the net magnetization $m$ of the entire nuclear spin bath for $I=1/2$, 1, and 3/2. This dependence comes from the Overhauser effect in which nuclear spin polarization acts as an effective magnetic field seen by the electron spin. We incorporate this effect by adding a magnetization-dependent shift to the precession frequency of the electron:
\begin{equation}\label{eq-Overhauser_shift}
w_i^j(m)=\mathrm{w}_i^j (\omega_{e}\rightarrow\omega_{e0}+mA),
\end{equation}
where $\omega_{e0}$ denotes the contribution to the precession frequency due purely to the external magnetic field, and where we use $w_i^j(m)$ to denote the rate to flip from state $i$ to state $j$ in the presence of nuclear spin magnetization $m$. For nuclei of spin $I$, we can express this magnetization in terms of occupation numbers, $N_\ell$, for each of the nuclear spin states:
\begin{equation} \label{eq-def-m}
m = \sum_{\ell=-I}^I \ell~N_\ell.
\end{equation}
In Fig.~\ref{fig-fr}, results for zero quadrupolar angle, $\theta=0$, are shown in the $I>1/2$ cases. Even though the quadrupolar coupling is nonzero, $\nu_Q>0$, only $\Delta m_I=\pm1$ transitions are permitted in this case because when $\theta=0$, the only effect of the quadrupolar interaction is to modify the energy splittings between nuclear spin levels, and so the selection rules are still determined solely by the HF interaction. We discuss the effect of nonzero $\theta$ below.

Several salient features are evident in Fig.~\ref{fig-fr}. First of all, the flip rates are strongly peaked at magnetization $m\approx-\omega_{e0}/A$. In the spin 1/2 case, the precise location of the peak is the value of $m$ at which the argument of the sine in Eq.~\eqref{eq:spin1/2fr} vanishes since the flip rates are essentially given by squared sinc functions. For low to moderate external magnetic field strengths and large $N$, the terms involving $\omega_n$ and $A^2$ can be neglected, leaving $m\approx-\omega_{e0}/A$. Similar statements hold for $I=1$ and $I=3/2$ in the absence of quadrupolar effects, as is clear from Eqs.~\eqref{eq:spin1fr} and \eqref{eq:spin3/2fr}. The fact that the flip rates are maximal at $m\approx-\omega_{e0}/A$ can be understood from energy conservation: At these values, the effective Zeeman energy of the electron is almost zero, and thus so is the energy mismatch between the electron and nucleus. This in turn reduces the energy penalty for flip-flops, accelerating the transfer of polarization. Conversely, the overall decay of the flip rates away from $m\approx\omega_{e0}/A$ is due to the HF interaction becoming inefficient at overcoming the large energy mismatch between the electronic and nuclear spin splittings.

It is also evident in Fig.~\ref{fig-fr} that the flip rates vanish periodically as a function of $m$. The periodicity is also controlled by the arguments of the sine functions in the flip rates. These zeros correspond to values of $\omega_e$ for which complete flip-flops between the electronic and nuclear spins occur---polarization is transferred back and forth between the electron and nucleus an integer number of times within a single drive period $T_R$. Because there is no net polarization transfer, the flip rate vanishes. For $I>1/2$, the locations of these zeros depend on which pair of adjacent spin levels we consider, although this dependence fades away in the large $N$ limit, where $A\to0$. In the next section, we show that these flip-rate zeros play a central role in the phenomenon of mode-locking. 

Each pair of flip rates describing transitions between the same two spin levels are almost equal [see panels (b), (d), (f) of Fig.~\ref{fig-fr}]. As can be seen from Eqs.~\eqref{eq:spin1/2fr}-\eqref{eq:spin3/2fr}, the differences of these flip rates are proportional to $S_{e,z}^{ss}(m)$, and this component of the electron steady state is suppressed near $m\approx-\omega_{e0}/A$ because it is proportional to $\omega_e$ (see Eq.~\eqref{eq-ss-e-SV}). This is a reflection of the fact that when $\omega_e=0$, the electron steady state becomes polarized along the optical pulse axis (the $x$ direction), where it is no longer affected by the pulses and is thus stabilized. In the figure, we see that this combination of accelerated flip-flops and the suppression of $S_{e,z}^{ss}(m)$ near $m\approx-\omega_{e0}/A$ results in flip rate differences that are more than two orders of magnitude smaller than the flip rates themselves.

The effect of a nonzero quadrupolar angle $\theta$ on the flip rates is shown in Figs.~\ref{fig-fr-one-nonzero-theta} and \ref{fig-fr-three-half-nonzero-theta} for $I=1$ and 3/2, respectively. In the case $I=1$, it is evident that $\theta$ has a negligible effect on the $\Delta m_I=\pm1$ flip rates. On the other hand, sufficiently large values of the angle, $\theta\gtrsim\pi/4$, give rise to $\Delta m_I=\pm2$ transitions that are not otherwise present. Although the rates for these transitions are two orders of magnitude smaller than those of the $\Delta m_I=\pm1$ transitions, they are still large enough to affect the polarization distribution of the nuclear spin bath, as we show in Sec.~\ref{sec:polarization_distr_feedback}. Similar but somewhat more prominent effects are evident for $I=3/2$ in Fig.~\ref{fig-fr-three-half-nonzero-theta}. Here, larger values of $\theta$ produce small but noticeable changes in $\Delta m_I=\pm1$ flip rates, significant $\Delta m_I=\pm2$ transition rates, and even $\Delta m_I=\pm3$ transitions. A striking feature evident in both Figs.~\ref{fig-fr-one-nonzero-theta} and \ref{fig-fr-three-half-nonzero-theta} is that the flip rates for $\Delta m_I=\pm2$ transitions do not decay as $m$ moves away from $m=-\omega_{e0}/A$. This is consistent with the fact that spin flips caused by the quadrupolar interaction do not require the electron and nuclear spin Zeeman energies to be equal. Unlike HF spin flips, quadrupolar spin flips depend weakly on the bath magnetization. On the other hand, the $\Delta m_I=\pm3$ flip rates are sensitive to $m$ (see Fig.~\ref{fig-fr-three-half-nonzero-theta}(b)), because these arise from a higher-order process that combines HF and quadrupolar spin flips.

%%%%%%%%%%%%%%%%%%%%%%%%%%%%%%%%%%

\subsection{Kinetic equations for multi-nuclear spin polarization distributions} \label{sec-ke}
In this section, we use the flip rates obtained in the previous section to construct kinetic rate equations that govern the evolution of the polarization distribution of the entire nuclear spin bath. We do this for each of the three values of nuclear total spin $I$ considered in this work. Although the kinetic equation for $I=1/2$ has been discussed in detail elsewhere \cite{BarnesPRL2011,EconomouPRB2014}, here we present an analytical solution to this equation that was not previously known. The kinetic equations for $I=1$ and 3/2 will be solved numerically in the next section to obtain nuclear spin polarization distributions in these cases. Detailed comparisons of the polarization distributions that result in all three cases for various parameter values are given below in Sec.~\ref{sec:polarization_distr_feedback}. In that section, these distributions are then used to compute the effect on the electron spin evolution with and without quadrupolar interactions.

%%%%%%%%%%%%%%%%%%%%%%

\subsubsection{Kinetic equation for spin $I=1/2$ nuclei} \label{sec-I-half}

The polarization of a spin 1/2 nuclear bath in a definite configuration with occupation numbers $N_+$ and $N_-$ (the number of spins in the $\ket{+1/2}$ and $\ket{-1/2}$ states, respectively) is given by $m=(N_+ - N_-)/2$. The total number of spins is $N=N_++N_-$. Knowledge of the polarization $m$ is sufficient to determine the two occupation numbers, $N_+$ and $N_-$. This in turn means that the probability of each bath configuration is equal to the polarization probability distribution $P(m)$.  We may write down a kinetic equation governing the dynamics of this distribution \cite{BarnesPRL2011,EconomouPRB2014}:
\begin{eqnarray}\label{eq-kinetic-half}
\frac{d}{dt}P(m)&=&-\sum_{\pm}\left[w_\pm(m){N\mp
2m\over2}\right]P(m) 
\\
&+&\sum_{\pm}w_\mp(m\pm1)\left[{N\pm
2m\over2}+1\right]P(m\pm1).\nonumber
\end{eqnarray}
A close look at this kinetic equation reveals that the right-hand side is comprised of two terms that are related to each other by shifting $m \to m+1$:
\begin{equation} \label{eq-P-half}
\frac{d}{dt}P(m) = F(m+1)-F(m),
\end{equation}
where $F(m)=w_-(m)(m+N/2)P(m)-w_+(m-1)(-m+1+N/2)P(m-1)$.
Therefore, in the steady state where $dP(m)/dt=0$, we find $F(m)=F(m+1)=\text{constant}$. Since we must have $P(N+1)=0$, it follows that this constant is zero. The equation $F(m)=0$ then yields a two-term recursion relation \citep{BarnesPRL2011,EconomouPRB2014}:
\begin{equation}\label{eq-rec-half}
P(m)= \frac{N-2m+2}{N+2m}~\frac{w_+(m-1)}{w_-(m)} P(m-1).
\end{equation}
This relation can easily be solved iteratively starting from an arbitrary value for $P(-N)$ and then imposing the normalization condition $\sum_mP(m)=1$. This approach was used to produce numerical results for the polarization distribution in Refs.~\cite{BarnesPRL2011,EconomouPRB2014}.

Here, we obtain an analytical solution for $P(m)$ by exploiting the explicit, non-perturbative expressions we obtained for the flip rates in Eq.~\eqref{eq:spin1/2fr}. First of all, an expression for $P(m)$ follows immediately from Eq.~\eqref{eq-rec-half}:
\begin{eqnarray}
    P(m)&=&\mathcal{N}^{-1}\prod_{k=1-N/2}^{m}\frac{N-2k+2}{N+2k}~\frac{w_+(k-1)}{w_-(k)}\nonumber\\
    &=&\frac{\mathcal{N}^{-1}N!}{(N/2+m)!(N/2-m)!}\prod_{k=1-N/2}^{m}\frac{w_+(k-1)}{w_-(k)},\nonumber\\
\end{eqnarray}
where $\mathcal{N}$ is a normalization factor. Next, we use the fact that the two flip rates only differ by the sign in front of $S_{e,z}^{ss}(m)$, which leads to a cascade of cancellations between the numerator and denominator in the product. We are left with
\begin{eqnarray}\label{eq:analyticSpinHalfPofm}
    P(m)&=&\frac{\mathcal{N}^{-1}}{(N/2+m)!(N/2-m)!}\prod_{k=1-N/2}^{m}\frac{1+S_{e,z}^{ss}(k-1)}{1-S_{e,z}^{ss}(k)}\nonumber\\&&\times\frac{(\omega_{e0}-\omega_n+Am)^2+A^2}{\sin^2(T_R\sqrt{(\omega_{e0}-\omega_n+Am)^2+A^2}/2)},
\end{eqnarray}
where we have absorbed additional constants into $\mathcal{N}$. The first, combinatoric factor in $P(m)$ corresponds to a Gaussian-like envelope that quickly approaches a Gaussian as $N$ increases: $[(N/2)!]^2/[(N/2+m)!(N/2-m)!]\to e^{-2m^2/N}$ as $N\to\infty$. The second factor in Eq.~\eqref{eq:analyticSpinHalfPofm} produces sharp spikes at values of $m$ that correspond to the zeros of the flip rates. These values of $m$ satisfy
\begin{equation}
    \sqrt{(\omega_{e0}-\omega_n+Am)^2+A^2}\approx\frac{2\pi p}{T_R},
\end{equation}
where $p$ is an integer.
The concentration of probability near these special values of $m$ produces mode-locking: Nuclear polarization shifts the electron Zeeman frequency to values where HF flip-flops stop transferring polarization between the electronic and nuclear spins. This happens because an integer number of flip-flops occur during each drive period. Using that $\omega_n\ll\omega_{e0}$ and assuming $N$ is sufficiently large that $A\ll\omega_{e0}$, these values of $m$ correspond to the electron precession becoming commensurate with the pulse train: $\omega_e=\omega_{e0}+Am\approx2\pi p/T_R$, which is the primary signature of mode-locking seen in experiments \cite{Greilich}.

The middle factor (the product) in Eq.~\eqref{eq:analyticSpinHalfPofm} is primarily responsible for the average magnetization of the nuclear spin bath, $\langle m\rangle=\sum_m mP(m)$. This factor is also where additional pulse parameters such as the rotation angle $\phi$ and the residual ground state population $q_0$ influence the polarization distribution. If $\phi$ is equal to 0 or $\pi$ or if $q_0$ is zero, then $S_{e,z}^{ss}(k)=0$ for all $k$, in which case the final factor in Eq.~\eqref{eq:analyticSpinHalfPofm} reduces to 1. In this case, the combinatoric factor, which is centered about $m=0$, ensures that the average magnetization will be small, $\langle m\rangle\approx 0$. On the other hand, if $\phi\ne0$ and the external magnetic field is sufficiently large, then $\langle m\rangle$ can be significant, and its sign depends on the sign of $\phi$ and on the orientation of the external field. If $\phi>0$, then $S_{e,z}^{ss}(m)$ is more often positive than negative for $m<-\omega_{e0}/A$, which in turn means that $\frac{1+S_{e,z}^{ss}(m-1)}{1-S_{e,z}^{ss}(m)}$ is biased toward values larger than 1, and so the product grows as $m$ increases. Once $m$ passes $-\omega_{e0}/A$, $S_{e,z}^{ss}(m)$ now tends to more negative values, and the product shrinks as $m$ increases. Thus, we see that for $\phi>0$, the product in Eq.~\eqref{eq:analyticSpinHalfPofm} is peaked at $m\approx-\omega_{e0}/A$, and so the average magnetization will lie between 0 and $-\omega_{e0}/A$. On the other hand, if $\phi<0$, then the same reasoning leads to the conclusion that the product in Eq.~\eqref{eq:analyticSpinHalfPofm} has a dip at $m\approx-\omega_{e0}/A$, and thus the net magnetization is driven away from this point and will have a sign that coincides with that of $\omega_{e0}$. These features are borne out in plots of Eq.~\eqref{eq:analyticSpinHalfPofm}, as shown below in Sec.~\ref{sec:polarization_distr_feedback}.

%%%%%%%%%%%%%%%%%%%%%%

\subsubsection{Kinetic equation for spin $I=1$ nuclei} \label{sec-I-one}

Before we write down the kinetic equation for $I=1$ nuclei, we first introduce the notation we use to distinguish different bath configurations.
We denote the occupation numbers of the three spin states by $N_{-1}$, $N_0$, and $N_1$. The bath polarization for a given configuration is then  $m=+1\times N_1 + 0\times N_0 -1 \times N_{-1}$. We see immediately that there is an important difference compared to the $I=1/2$ case considered above: The polarization does not uniquely specify a configuration of the bath. For instance, in the case of two $I=1$ spins with $m=0$, we can have either $N_1=1=N_{-1}$ and $N_0=0$ or $N_1=0=N_{-1}$ and $N_0=2$. This is in contrast to the $I=1/2$, where each value of $m$ corresponds to a unique configuration. As the number of spins increases, the number and orders of such ``degeneracies" grow quickly. Because the polarization does not uniquely specify a configuration, we must combine it with one of the occupation numbers to uniquely label different configurations. We choose to use $N_0$ and express the probability of a given configuration by $P(m,N_0)$. Unlike in the spin 1/2 case, this quantity is now distinct from the polarization probability distribution; the latter is obtained by summing over all possible values of $N_0$ that are consistent with the given value of $m$:
\begin{equation} \label{eq-PmN0}
P(m) = \sum_{N_0} P(m,N_0).
\end{equation}

We can write down a kinetic equation for $P(m,N_0)$:
\begin{eqnarray}\label{eq-kinetic-one}
\frac{d}{dt}P(m,N_0) &=& F(m,N_0) + G(m+1,N_0 -1) \nonumber\\
&&-G(m,N_0) - F(m+1,N_0+1),
\end{eqnarray} 
where
\begin{eqnarray}
&&F(m,N_0) = -w_{0}^{-1} P(m,N_0) N_0 \\\label{eq-F}&&+ w_{-1}^0 (m-1) P(m-1,N_0-1) N_- (m-1,N_0-1),\nonumber\\
&&G(m,N_0) = w_{1}^{0} P(m,N_0) N_+(m,N_0)\nonumber\\&&-w_{0}^1 (m-1) P(m-1,N_0+1) (N_0 +1).\label{eq-G}
\end{eqnarray}
Here $N_\pm(m,N_0) \equiv (1/2)(N\pm m-N_0)$. In the kinetic equation above we have only considered the $\Delta m_I=\pm 1$ transitions. Including transitions that change the angular momentum by more than 1 (for instance due to quadrupolar interactions) leads to additional terms not shown above. Such terms are illustrated for the case of $I=3/2$ nuclei in the next section. Returning to the spin 1 case, the steady state of the above kinetic equation,
\begin{equation} \label{eq-one-ss}
F(m,N_0) -G(m,N_0) =  F(m+1,N_0+1)-G(m+1,N_0 -1),
\end{equation}
does not yield a recursion relation as in the $I=1/2$ case. We solve this equation (and its generalization for nonzero quadrupolar interactions) numerically in Sec.~\ref{sec:polarization_distr_feedback}.

%%%%%%%%%%%%%%%%%%%%%%

\subsubsection{Kinetic equation for spin $I=3/2$ nuclei} \label{sec-I-3halves}

We again adopt the notation $\{ ++ ,+,-,--\}$ to label quantities associated with the four spin quantum numbers $m_I=\{+3/2,+1/2,-1/2,-3/2 \}$ of a spin 3/2 nucleus. For a nuclear spin bath comprised of $N=N_\pp+N_\p + N_\m + N_\mm$ spins, the magnetization of the system (Eq.~(\ref{eq-def-m})) is $m= (3N_\pp+N_\p - N_\m -3 N_\mm)/2$. In the $I=3/2$ case, we need two more quantities in addition to $m$ to uniquely label different multi-spin configurations. We choose these to be $N_{++}$ and $N_{--}$. The remaining two occupation numbers are then determined by these three quantities for a fixed total number of spins:
\begin{eqnarray}
N_\p &=& \frac{1}{2}(2m+N-4 N_\pp+2N_\mm ),\\
 N_\m &=& \frac{1}{2}(-2m+N +2 N_\pp-4N_\mm).
\end{eqnarray}
The probabilities $P(m,N_\pp,N_\mm)$ that the nuclear spin bath is in the various configurations labeled by $m$, $N_\pp$, and $N_\mm$ obey the following set of kinetic equations:
\begin{widetext}

\begin{eqnarray}\label{eq-kinetic-threehalves}
\frac{d}{dt}P(m,N_\pp,N_\mm) &=& F(m,N_\pp,N_\mm)+G(m,N_\pp,N_\mm) + H(m,N_\pp,N_\mm)  \nonumber\\
&&+  I(m,N_\pp,N_\mm) + J(m,N_\pp,N_\mm) \nonumber \\
&&-  F(m+1,N_\pp+1,N_\mm) - G(m+1,N_\pp,N_\mm-1) - H(m+1,N_\pp,N_\mm)  \nonumber \\
&&-  I(m-2,N_\pp-1,N_\mm) - J(m+2,N_\pp,N_\mm-1),
\end{eqnarray} 

where

\begin{eqnarray}
F(m,N_\pp,N_\mm)  &=&   + w_\p^\pp(m-1) P (m-1,N_\pp -1,N_\mm)  N_\p(m-1,N_\pp -1 ,N_\mm)\nonumber\\&&-w_\pp^\p (m) P(m,N_\pp, N_\mm) N_\pp,\\
 G(m,N_\pp,N_\mm)&=&  +  w_\mm^\m(m-1) P (m-1,N_\pp ,N_\mm+1)(N_\mm +1) \nonumber \\
  &&- w_\m^\mm(m) P(m,N_\pp, N_\mm) N_\m(m,N_\pp ,N_\mm),\\
  H(m,N_\pp,N_\mm)&=&  +  w_\m^\p(m-1) P (m-1,N_\pp ,N_\mm)N_\m(m-1,N_\pp ,N_\mm)\nonumber\\&&-w_\p^\m(m) P(m,N_\pp, N_\mm) N_\p(m,N_\pp ,N_\mm),\\
  I(m,N_\pp,N_\mm)  &=& + w_\pp^\m(m+2) P (m+2,N_\pp +1,N_\mm)  (N_\pp+1)\nonumber\\&& -w_\m^\pp (m) P(m,N_\pp, N_\mm) N_\m (m,N_\pp,N_\mm),\\
 J(m,N_\pp,N_\mm)  &=& +  w_\mm^\p(m-2) P (m-2,N_\pp ,N_\mm+1)  (N_\mm+1)\nonumber\\&&-w_\p^\mm (m) P(m,N_\pp, N_\mm) N_\p (m,N_\pp,N_\mm).
\end{eqnarray}

\end{widetext}

Here, we have included $\Delta m_I=\pm1$ and $\Delta m_I=\pm2$ transitions. Although $\Delta m_I=\pm3$ transitions cannot be directly driven by either the HF interaction or the quadrupolar interaction to first order in their respective coupling strengths, they can potentially arise from higher-order effects as we saw from the flip rates in Fig.~\ref{fig-fr-three-half-nonzero-theta}. Now that we have the kinetic equations governing the nuclear polarization, the next step is to solve them.

%%%%%%%%%%%%%%%%%%%%%%%%%%%%%%%%%%%%%%%%%%%%%%%%%%
%%%%%%%%%%%%%%%%%%%%%%%%%%%%%%%%%%%%%%%%%%%%%%%%%%
%%%%%%%%%%%%%%%%%%%%%%%%%%%%%%%%%%%%%%%%%%%%%%%%%%
%%%%%%%%%%%%%%%%%%%%%%%%%%%%%%%%%%%%%%%%%%%%%%%%%%

\section{Nuclear polarization distribution and feedback}\label{sec:polarization_distr_feedback}

\subsection{Steady-state polarization distributions}\label{sec:polarization_dist}

\begin{figure}
\includegraphics[scale=.4]{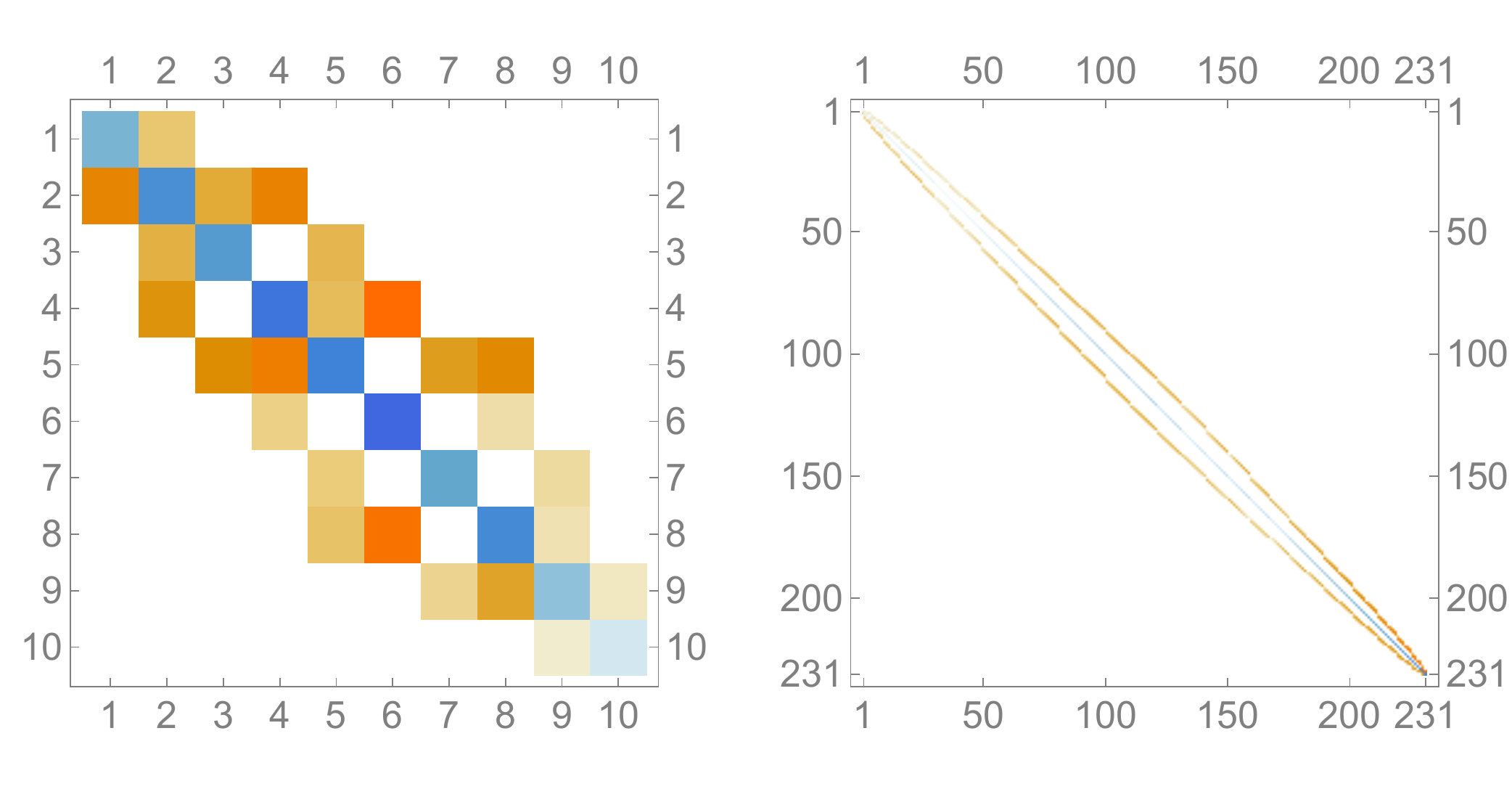}
\caption{Structure of the matrix $\cal R$ defining the linear system of equations governing the steady-state solution of the multi-nuclear kinetic equation for spin $I=1$ for (left) $N=3$ spins and (right) $N=20$ spins in the absence of quadrupolar interactions.}
\label{fig-sparse}
\end{figure}

\begin{figure*}
\includegraphics[scale=.67]{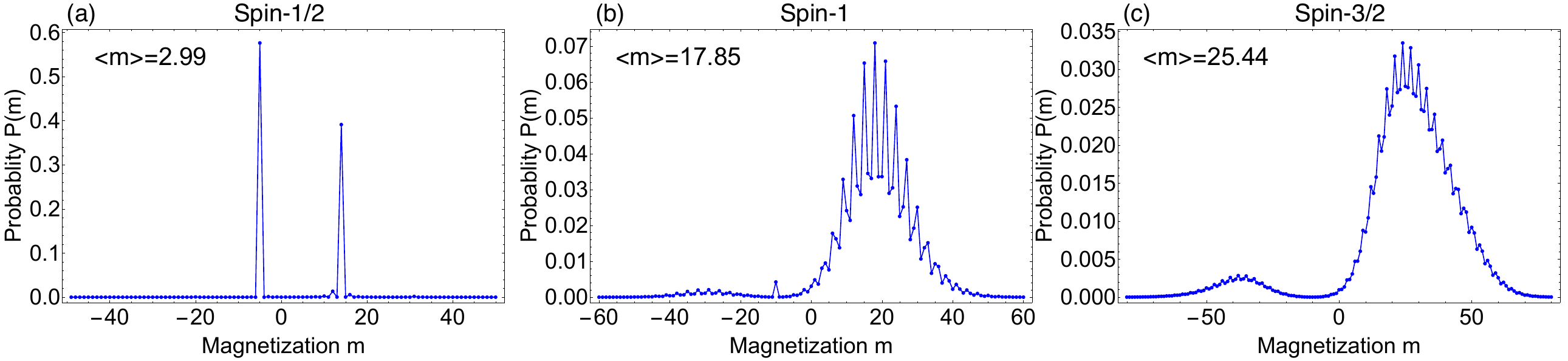}
\caption{Steady-state nuclear spin polarization distribution of a bath with $N=200$ nuclear spins for (a) $I=1/2$, (b) $I=1$, and (c) $I=3/2$. The parameter values are
$T_R=13.2$~ns, $NA=10$~GHz, $\omega_{e0}=0.5$~GHz, $\omega_n=-0.5$~MHz,  $\gamma_e=0.5$~GHz, $q_0=0.3$,
$\phi=-\pi/2$, $\nu_Q = 2.8$~MHz. In the case of $I=3/2$ and $I=1$ the quadrupolar angle is $\theta=0$. }
\label{fig-200-pol-compare}
\end{figure*}

\begin{figure}
\includegraphics[scale=.295]{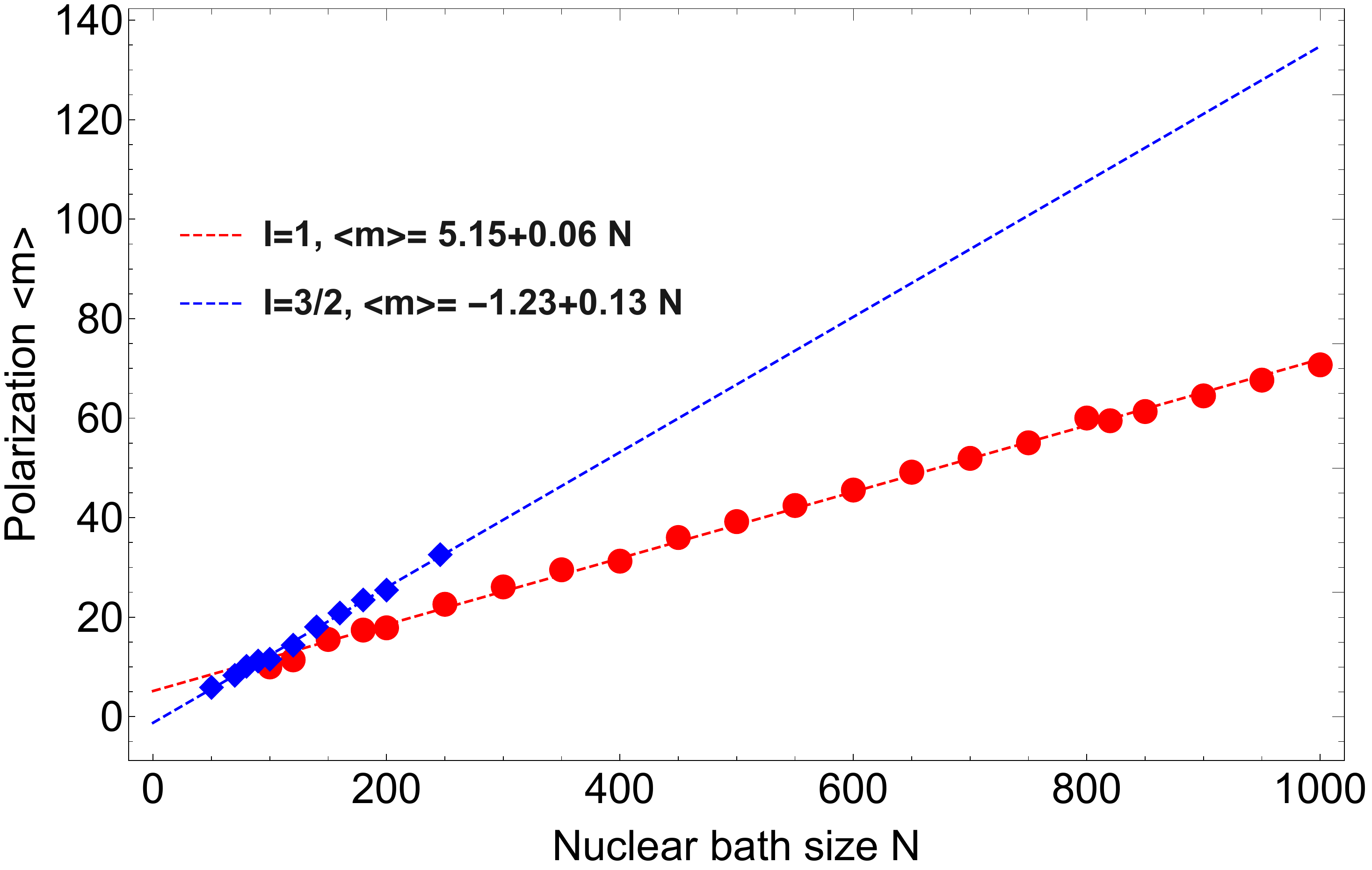}
\caption{Extrapolation of the average nuclear spin bath polarization $\langle m\rangle$ to larger bath sizes $N$ for two values of total spin: $I=1$ (red circles) and $I=3/2$ (blue diamonds). The points are obtained by solving the respective kinetic equations, Eqs.~\eqref{eq-kinetic-one} and \eqref{eq-kinetic-threehalves}. The lines are linear fits. The parameter values are
$T_R=13.2$~ns, $NA=10$~GHz, $\omega_{e0}=0.5$~GHz, $\omega_n=-0.5$~MHz,  $\gamma_e=0.5$~GHz, $q_0=0.3$,
$\phi=-\pi/2$, $\nu_Q = 2.8$~MHz, $\theta=0$. }
\label{fig-extrap}
\end{figure}

\begin{figure*}
\includegraphics[scale=.66]{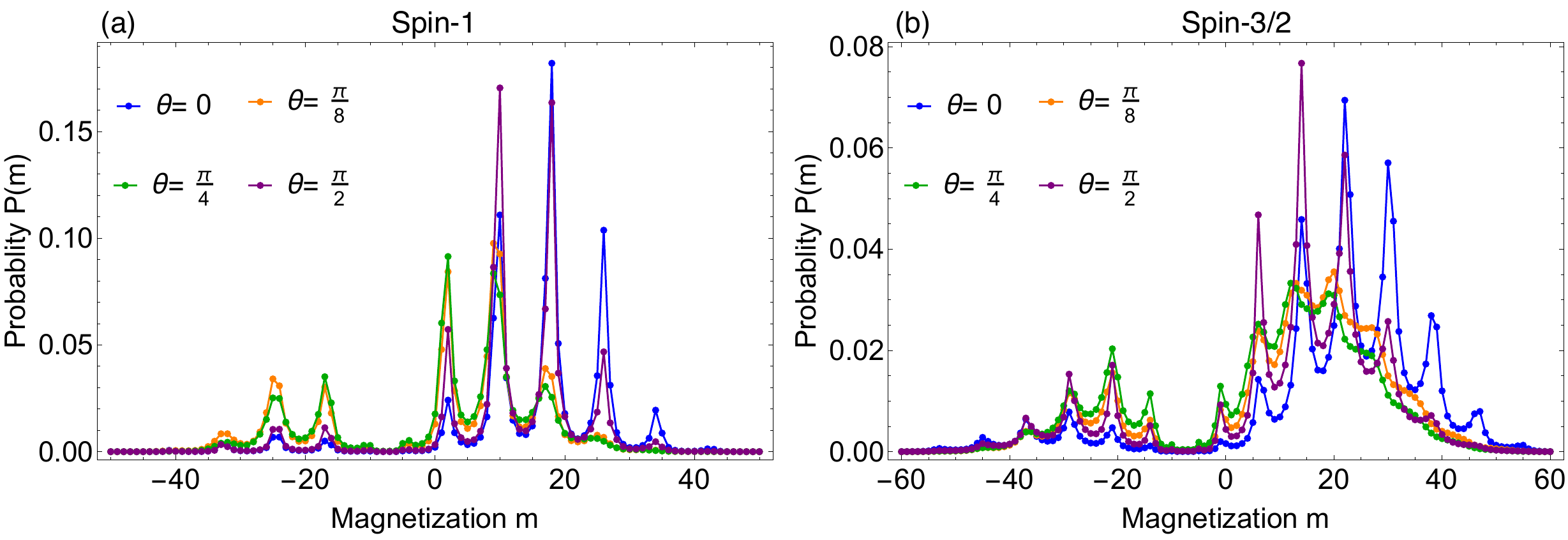}
\caption{Steady-state nuclear spin polarization distribution of a bath with $N=150$ nuclear spins for four different values of the quadrupolar angle $\theta$ for (a) $I=1$ and (b) $I=3/2$. The parameter values are
$T_R=13.2$~ns, $NA=10$~GHz, $\omega_{e0}=0.5$~GHz, $\omega_n=-0.5$~MHz,  $\gamma_e=0.5$~GHz, $q_0=0.3$,
$\phi=-\pi/2$, $\nu_Q = 2.8$~MHz.}
\label{fig-150-pol-comparison}
\end{figure*}

For $I=1$ and $I=3/2$, we solve the respective kinetic equations numerically to obtain steady-state polarization distributions. This is done by first setting the time derivatives to zero: $\frac{d}{dt}P(m,N_\pp,N_\mm)=0$. The resulting algebraic equations are then collected together and written as a matrix $\cal R$ acting on a vector $\cal V$ of the probabilities $P(m,N_\pp,N_\mm)$ such that ${\cal R}{\cal V}=0$. Thus, the steady-state polarization distribution is the unique null vector of $\cal R$. The matrix $\cal R$ depends on the Overhauser-shifted flip rates and occupation numbers for each configuration. The linear dimension of this matrix is equal to the number of distinct multi-spin configurations. For $N$ spins of total spin $I$, the number of configurations is given by the simplicial polytopic numbers $\binom{N+2I}{2I}$. For $I=1/2$, 1, and 3/2, this gives $N+1$, $(N+1)(N+2)/2$, and $(N+1)(N+2)(N+3)/6$, respectively. Therefore, in the case of $I=1$, we must compute the null vector of a matrix that grows quadratically with the number of nuclei, while for $I=3/2$, we must do the same for a matrix that grows like $N^3$. The matrix $\cal R$ is quite sparse in both cases (see Fig.~\ref{fig-sparse}), especially in the absence of quadrupolar interactions. This allows us to employ the Arnoldi method to compute the steady-state polarization distribution for hundreds of spins with $I=3/2$ and thousands of spins with $I=1$.

Fig.~\ref{fig-200-pol-compare} compares results for the steady-state nuclear spin polarization for $N=200$ for all three values of $I$. In the $I>1/2$ cases, we set the quadrupolar angle to zero, $\theta=0$; however, the nonzero quadrupolar interaction $\nu_Q>0$ still modifies the energy splittings between the nuclear spin levels. In all three cases, the polarization distribution exhibits multiple narrow peaks at values of $m$ that correspond to the mode-locking frequencies, i.e., these values of $m$ are such that $\omega_{e0}+Am = 2\pi p/T_R$ where $p$ is an integer (for an analytical derivation of the $I=1/2$ case see Section~\ref{sec-I-half}). As discussed in Sec.~\ref{sec-fliprates}, the flip rates approximately vanish at these values of $m$. (Note that the spacing of the peaks in Fig.~\ref{fig-200-pol-compare} is five times smaller than the spacing of the flip-rate zeros in Figs.~\ref{fig-fr}, \ref{fig-fr-one-nonzero-theta}, and \ref{fig-fr-three-half-nonzero-theta} because this spacing is proportional to $1/A=N/{\cal A}$, and $N$ is five times smaller in Fig.~\ref{fig-200-pol-compare}.) The steady-state probabilities $P(m,N_\pp,N_\mm)$ are largest at these magnetization values because they are multiplied by nearly vanishing flip rates in the kinetic equations; the probabilities must compensate for the smallness of the flip rates such that the product of the two is finite and comparable to terms of similar size in the kinetic equations. This trend can be seen explicitly from the analytical solution in the $I=1/2$ case, Eq.~\eqref{eq:analyticSpinHalfPofm}, where it is evident that $P(m)$ depends inversely on the flip rates. In Fig.~\ref{fig-200-pol-compare}, we see that this also occurs for $I>1/2$. For all values of $I$, we can physically understand the formation of probability peaks at flip-rate zeros as resulting from the fact that, at these magnetization values, the joint electron-nuclear spin evolution under the HF interaction becomes commensurate with the driving pulses. Consequently, the pulses do not cause a net polarization transfer between the electron and nuclear spins. Thus, these values of the magnetization $m$ provide a point of stability in the electron-nuclear feedback mechanism. We also see from Fig.~\ref{fig-200-pol-compare}(a), and to some degree from Fig.~\ref{fig-200-pol-compare}(b), that the polarization distribution is suppressed in the vicinity of $m=-\omega_{e0}/A$ (which corresponds to $m=-10$ for the parameters used in the figure). This is due to the fact that the flip rates are largest near these magnetization values and therefore drive population away from these values.

Another striking feature of the polarization distributions in Fig.~\ref{fig-200-pol-compare} is that the distributions for $I>1/2$ exhibit broad envelopes in addition to the mode-locking peaks. This is a consequence of the fact that there are multiple distinct flip rates for $I>1/2$, as shown in Eqs.~\eqref{eq:spin1fr} and \eqref{eq:spin3/2fr}. These flip rates oscillate with $\omega_e$ at distinct frequencies that differ from each other by an amount proportional to $A$. Therefore, they do not all vanish at the same values of $\omega_e$, dulling the sharpness of the mode-locking peaks. This effect becomes diminished at larger $N$, because in this limit $A$ decreases, and all the flip-rate zeros approach the values of $m$ at which $\omega_{e0}+Am=2\pi p/T_R$, where $p$ is an integer, producing a more comb-like distribution. The broadening of the distribution at smaller values of $N$ is an important feature that is missed when $I=1/2$ spins are used to model $I>1/2$ spin baths. In the example of Fig.~\ref{fig-200-pol-compare}, we see that it also leads to an increase in the average magnetization $\langle m\rangle$ due to the enhanced weight of the distribution at positive magnetizations. This enhancement is more pronounced for $I=3/2$ compared to $I=1$. Fig.~\ref{fig-extrap} examines the behavior of $\langle m\rangle$ as a function of $N$. The points are obtained by solving the respective kinetic equations, Eqs.~\eqref{eq-kinetic-one} and \eqref{eq-kinetic-threehalves}. In the $I=1$ case, it is possible to obtain results for much larger bath sizes because the $\cal R$ matrix is much smaller in this case. For both $I=1$ and $I=3/2$, the points are well described by a linear relationship between $\langle m\rangle$ and $N$, as shown in the figure. We find that for the parameters considered and for large $N$, the average polarization for $I=3/2$ is approximately two times larger compared to that of an $I=1$ bath, with the net polarization in this case approaching 9\%.

The effects of nonzero quadrupolar angle on the polarization distribution for $I=1$, 3/2 are illustrated in Fig.~\ref{fig-150-pol-comparison}. Here, we set $N=150$, because nonzero $\theta$ reduces the sparsity of the $\cal R$ matrix, making the numerical computation more intensive than before, especially for $I=3/2$. From Fig.~\ref{fig-150-pol-comparison}(a), we see that for $I=1$, nonzero $\theta$ leads to quantitative changes in the heights of the mode-locking peaks, along with a slight redistribution of the probability to negative magnetizations for intermediate values of $\theta$. Similar behavior occurs for $I=3/2$, as shown in Fig.~\ref{fig-150-pol-comparison}(b). The redistribution can be understood from the fact that, in the absence of the HF interaction, the quadrupolar coupling produces a Gaussian distribution centered around $m=0$. This is discussed in more detail below. The fact that this redistribution is strongest near $\theta=\pi/4$ suggests that the $\Delta m_I=\pm1$ quadrupolar-driven transitions play an important role in this process. This effect constitutes another way in which the quadrupolar interaction can make the DNP process for $I>1/2$ depart significantly from what is predicted for an $I=1/2$ bath. Also notice that in both panels of Fig.~\ref{fig-150-pol-comparison}, the polarization distributions are still suppressed near $m=-\omega_{e0}/A$ even for $\theta>0$. This indicates that the HF contributions to the flip rates remain an important factor in shaping the overall distribution.

\begin{figure}
\includegraphics[scale=.35]{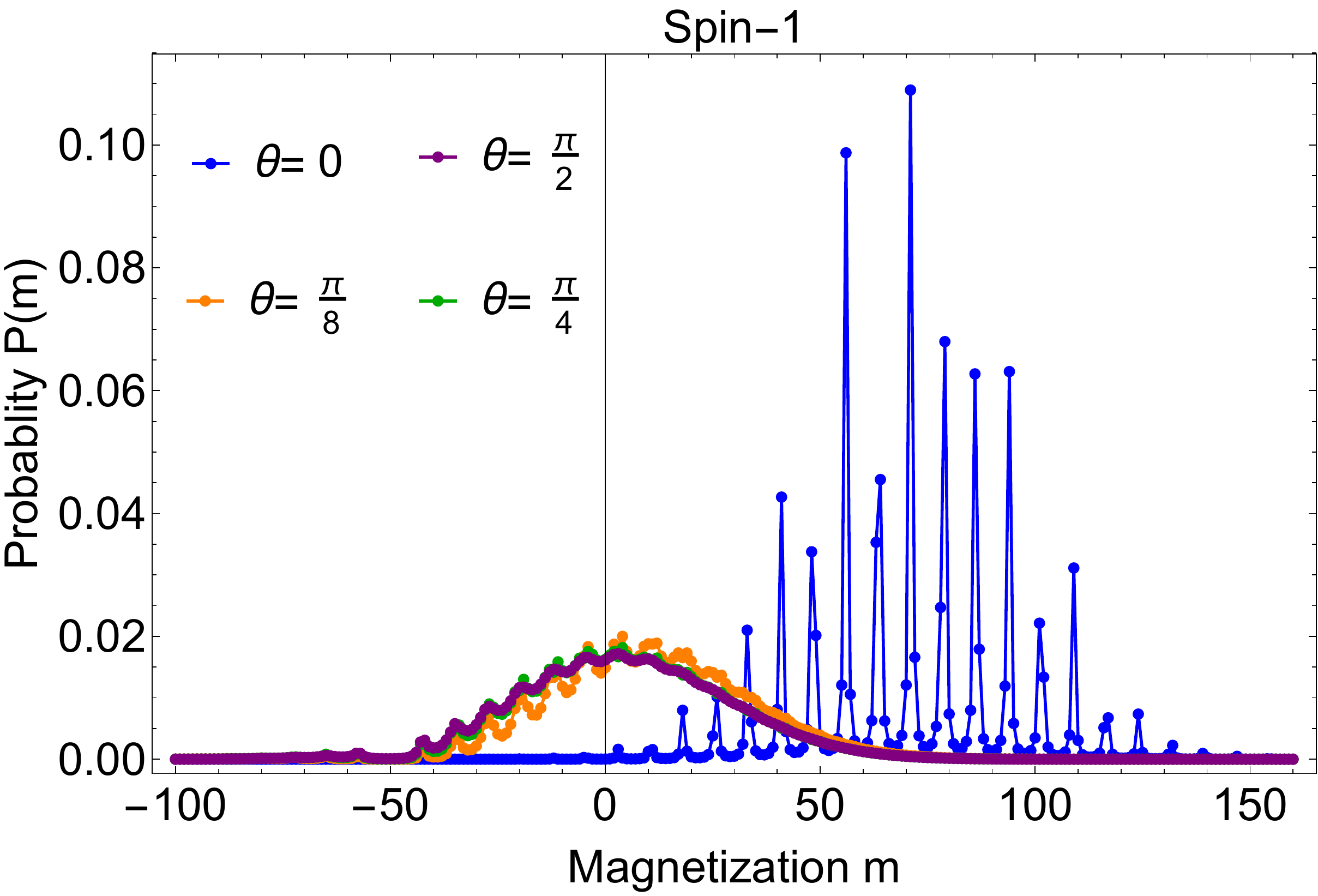}
\caption{Steady-state nuclear spin polarization distribution of a bath with $N=1000$ $I=1$ nuclear spins for four different values of the quadrupolar angle $\theta$. The other parameter values are
$T_R=13.2$~ns, $NA=10$~GHz, $\omega_{e0}=0.5$~GHz, $\omega_n=-0.5$~MHz,  $\gamma_e=0.5$~GHz, $q_0=0.3$,
$\phi=-\pi/2$, $\nu_Q = 2.8$~MHz.}
\label{fig-spin-one-pol-1000h}
\end{figure}

\begin{figure}
\includegraphics[scale=.28]{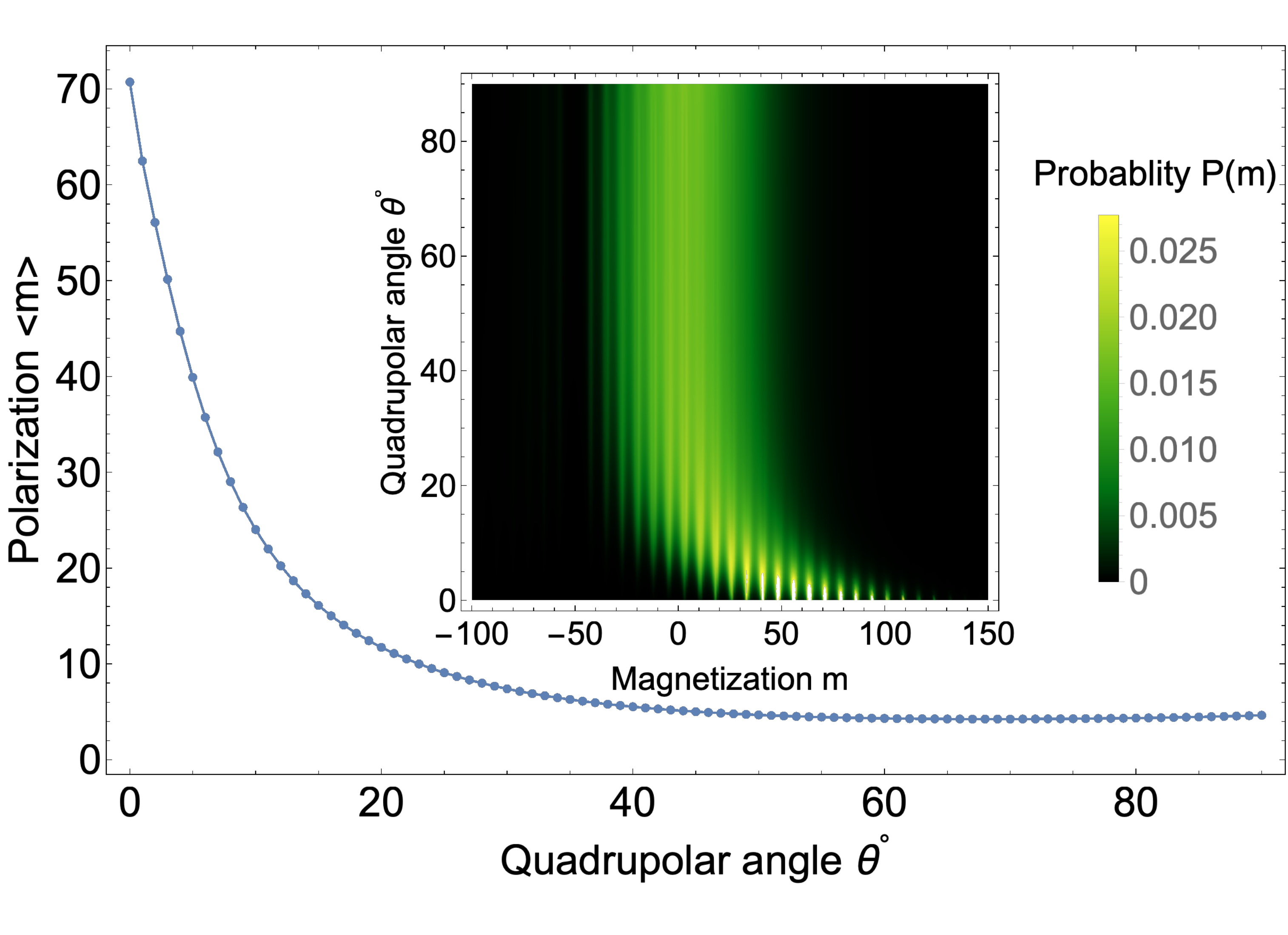}
\caption{The average polarization $\langle m\rangle$ of a nuclear spin bath with $N=1000$ nuclei of total spin $I=1$ for several values of the quadrupolar angle in the range of $0\le\theta\le\pi/2$. The inset color map shows the steady-state nuclear spin polarization distribution over the same range of quadrupolar angles. The other parameter values are
$T_R=13.2$~ns, $NA=10$~GHz, $\omega_{e0}=0.5$~GHz, $\omega_n=-0.5$~MHz,  $\gamma_e=0.5$~GHz, $q_0=0.3$,
$\phi=-\pi/2$, $\nu_Q = 2.8$~MHz.}
\label{fig-quad-angle}
\end{figure}

\begin{figure*}
\includegraphics[scale=.7]{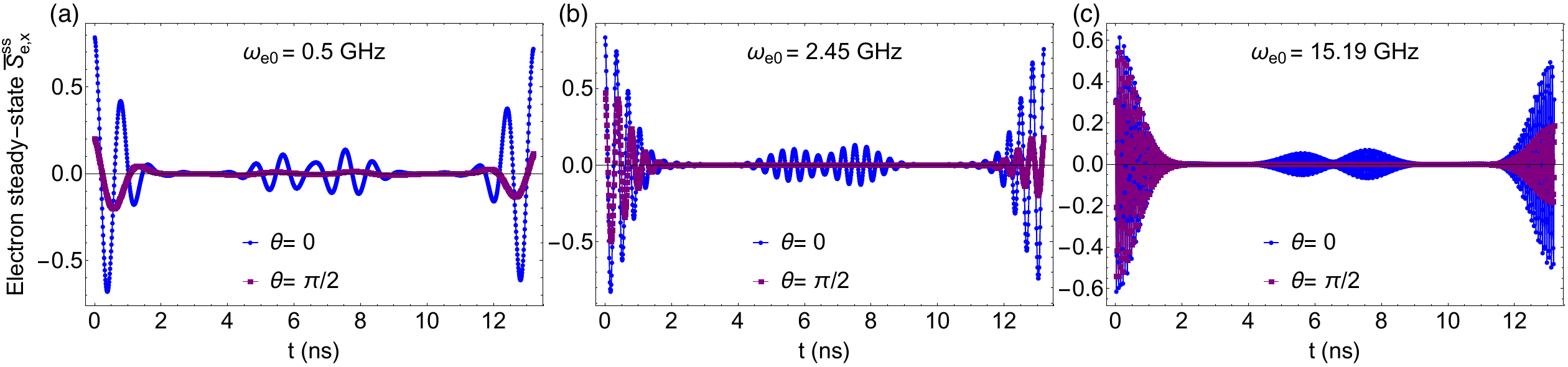}
\caption{The feedback effect of $N=1000$ $I=1$ nuclear spins on the $x$ component of electron spin steady state as a function of time over one drive period $T_R=13.2$~ns. Here the quadrupolar angles $\theta=0$ and $\theta=\pi/2$ are considered for different bare electron Zeeman frequencies of (a) 0.5 GHz, (b) 2.45 GHz and (c) 15.19 GHz. The electron Zeeman frequencies chosen for (b) and (c) correspond to the local minima shown in Fig.~\ref{fig-feedback-minima} and the nuclear spin polarization distribution for (a) is shown in Fig.~\ref{fig-spin-one-pol-1000h}. The parameter values are $NA=10$~GHz, $\omega_n=-0.5$~MHz,  $\gamma_e=0.5$~GHz, $q_0=0.3$,
$\phi=-\pi/2$, $\nu_Q = 2.8$~MHz. }
\label{fig-ss-feedback}
\end{figure*}

Fig.~\ref{fig-spin-one-pol-1000h} again shows the effect of nonzero $\theta$ for $I=1$, but now for a bath of size $N=1000$. For $\theta=0$, there is a distinct comb-like structure that is the hallmark of mode-locking. However, for $\theta>0$, this structure quickly disappears and is replaced by an almost Gaussian distribution centered around zero magnetization. A Gaussian distribution is in fact what occurs in the absence of the HF interaction, because the flip rates are then purely due to the quadrupolar coupling, which means that they are independent of $m$ and are equal for $\Delta m_I>0$ and $\Delta m_I<0$. This shows that the quadrupolar interaction plays a much more important role compared to the HF interaction for the case considered in Fig.~\ref{fig-spin-one-pol-1000h}. This is because the larger value of $N$ corresponds to a reduction in the HF coupling $A$, and hence in the magnitude of the flip rates (see Eq.~\eqref{eq:spin1fr}). This in turn increases the relative importance of the quadrupolar interaction. This can be seen from Fig.~\ref{fig-fr-one-nonzero-theta}, where it is evident that as $\theta$ increases, the flip rate for the $\Delta m_I=2$ transition quickly surpasses the difference in the flip rates for the $\Delta m_I=\pm1$ transitions. As a consequence, the probability distribution is no longer sensitive to the detailed features of the $\Delta m_I=\pm1$ transitions, which are responsible for both the comb-like mode-locking structure and the suppression near $m=-\omega_{e0}/A$. This shows that even small values of $\theta$ can have a dramatic effect on the DNP process for large numbers of nuclei. This is quantified in Fig.~\ref{fig-quad-angle}, which shows how the nuclear spin polarization distribution and average magnetization, $\langle m\rangle$, depend on $\theta$. The latter quickly decays with increasing $\theta$. As is evident from the inset in Fig.~\ref{fig-quad-angle}, the distribution itself exhibits mode-locking fringes at small $\theta$ that become blurred at larger $\theta$. The sensitivity of mode-locking to the quadrupolar interaction suggests that it could be used as a diagnostic tool to estimate the size of the quadrupolar coupling strength and angle in experiments. This is further supported in the next section, where we show how the steady-state electron spin vector in the presence of DNP feedback depends on the quadrupolar angle.

%%%%%%%%%%%%%%%%%%%%%%%%%%%%%%%%%%%%%%%%%%%%%%%%%%
%%%%%%%%%%%%%%%%%%%%%%%%%%%%%%%%%%%%%%%%%%%%%%%%%%
%%%%%%%%%%%%%%%%%%%%%%%%%%%%%%%%%%%%%%%%%%%%%%%%%%
%%%%%%%%%%%%%%%%%%%%%%%%%%%%%%%%%%%%%%%%%%%%%%%%%%

\subsection{Feedback on electron spin} \label{sec-feedback}

\begin{figure}
\includegraphics[scale=.43]{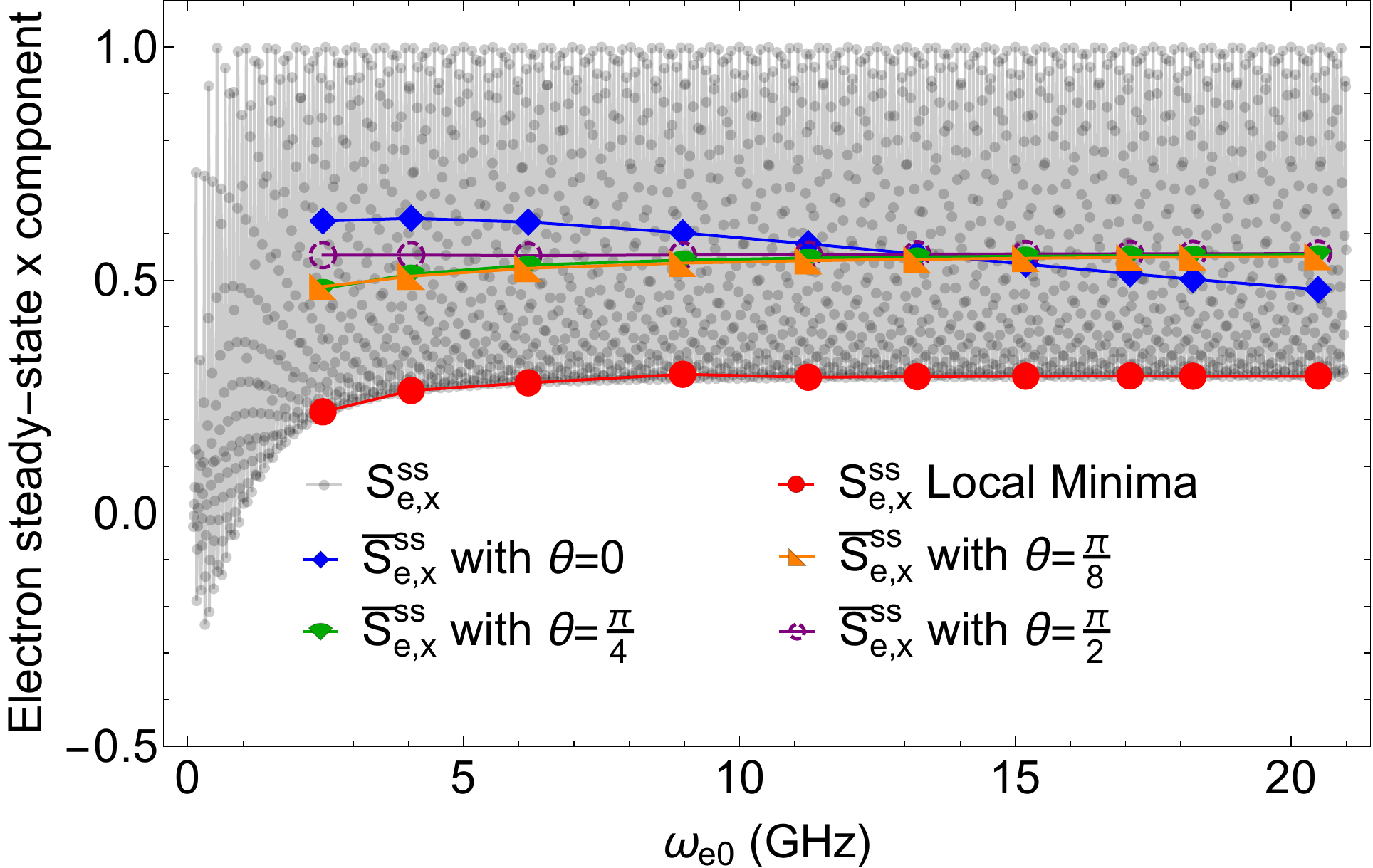}
\caption{The effect of the $I=3/2$ nuclear feedback on the $x$ component of the steady-state electron spin vector. The red filled circles indicate local minima of $S_{e,x}^{ss}$ (shown in gray) for several values of the electron Zeeman frequency $\omega_{e0}$ without nuclear feedback. The other points indicate the values of $\overline{S}_{e,x}^{ss}(\omega_{e0})$ at the same values of $\omega_{e0}$, but now with feedback included as in Eq.~\eqref{eq-feedback}. Results for four different values of the quadrupolar angle $\theta$ are shown. Other parameter values are $N=150$, $T_R=13.2$~ns, $NA=10$~GHz, $\omega_n=-0.5$~MHz,  $\gamma_e=0.5$~GHz, $q_0=0.3$,
$\phi=-\pi/2$, $\nu_Q = 2.8$~MHz.} 
\label{fig-feedback-minima}
\end{figure}

\begin{figure*}
\includegraphics[scale=.75]{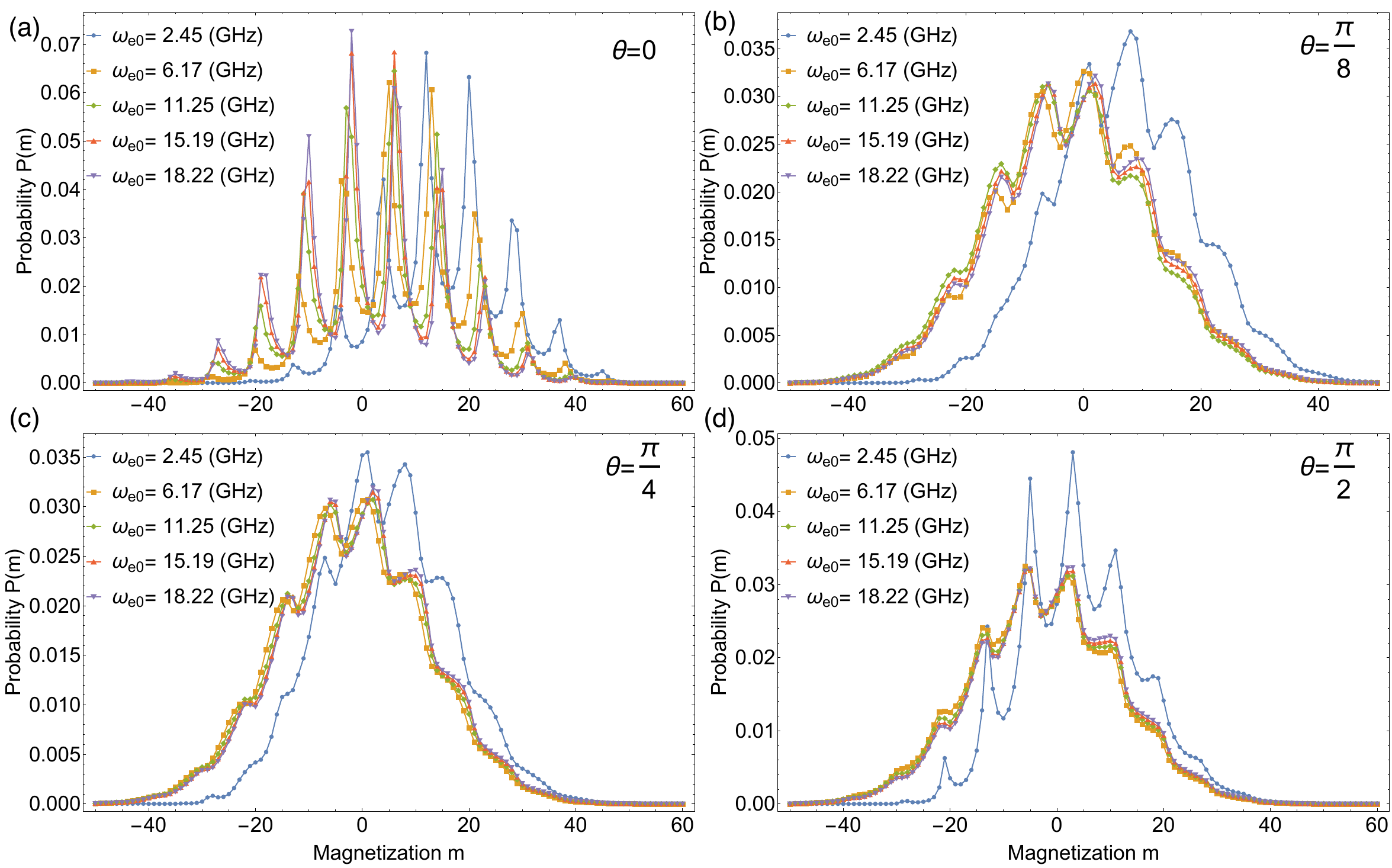}
\caption{The nuclear spin polarization distributions corresponding to five of the electron Zeeman frequency values from Fig.~\ref{fig-feedback-minima} for quadrupolar angles (a) $\theta=0$, (b) $\theta=\pi/8$, (c) $\theta=\pi/4$, and (d) $\theta=\pi/2$ for an $I=3/2$ nuclear bath. Other parameter values are $N=150$, $T_R=13.2$~ns, $NA=10$~GHz, $\omega_n=-0.5$~MHz,  $\gamma_e=0.5$~GHz, $q_0=0.3$,
$\phi=-\pi/2$, $\nu_Q = 2.8$~MHz. }
\label{fig-pols-omega-e}
\end{figure*}

Once we obtain the steady-state polarization distribution of the nuclear spin bath, the final step is to update the steady state of the electron by applying the Overhauser shift to the Zeeman frequency:
\begin{eqnarray} \label{eq-feedback}
\overline{S}^{ss}_{e,i}(t,\omega_{e0}) = \sum_{m}  P(m) S^{ss}_{e,i}(t,\omega_{e0}+mA).
\end{eqnarray}
Here the summation is over all possible values of $m$, and $t$ is the time elapsed since the last pulse. We obtain the time-evolved electron steady state by starting from the expression for the steady state immediately after a pulse, Eq.~\eqref{eq-ss-e-SV}, and evolving it under Larmor precession with frequency $\omega_{e0}+mA$ for time $t$. Fig.~\ref{fig-ss-feedback} shows the resulting DNP-modified electron steady state over one drive period for six different $N=1000$, $I=1$ polarization distributions. Two of these are distributions shown in Fig.~\ref{fig-spin-one-pol-1000h}---the ones corresponding to $\theta=0$ and $\theta=\pi/2$. The modified steady states for these two cases are shown in Fig.~\ref{fig-ss-feedback}(a), where it is evident that a large quadrupolar angle suppresses oscillations, both in the vicinity of the driving pulses and in the ``echo" that occurs in the middle of the drive period near $t=T_R/2$, which is 6.6 ns for the chosen parameter values. Similar behavior occurs for other values of the external magnetic field, as is demonstrated in Figs.~\ref{fig-ss-feedback}(b), (c). It should be noted that the amplitude of these oscillations are used to identify the presence of mode-locking \cite{Greilich}, and so the suppression of these oscillations can provide an experimental indicator of substantial quadrupolar effects.

The electron steady state, Eq.~\eqref{eq-ss-e-SV}, is a rapidly oscillatory function of the applied magnetic field. In Ref.~\cite{EconomouPRB2014}, it was found using perturbation theory that for $I=1/2$, nuclear feedback suppresses the amplitudes of these oscillations. In particular, it was shown that the $x$ component of the electron steady-state SV approaches unity for all values of the external magnetic field as a consequence of mode-locking: The SV becomes synchronized with the pulses such that it lies parallel to the optical axis at the pulse times. Here, we examine how this effect is modified by the presence of quadrupolar interactions. This is illustrated in the case of $I=3/2$ in Fig.~\ref{fig-feedback-minima}, where we show the $x$-component of the electron steady state immediately after a pulse, $S_{e,x}^{ss}$, for ten different values of the electron Zeeman frequency with and without feedback. We are primarily interested in the amplitude of the electron steady-state oscillations, so we choose the ten different Zeeman frequencies that correspond to minima of the oscillations in the absence of feedback (red dots in Fig.~\ref{fig-feedback-minima}). To find how the envelope of the electron spin oscillations is affected by the feedback process, we compute the nuclear spin polarization distributions for each of these minima. These distributions then alter the values of these minima according to Eq.~\eqref{eq-feedback} (with $t=0$). As can be seen from Fig.~\ref{fig-feedback-minima}, the amplitude of the electron steady-state oscillations is suppressed (i.e., the minima increase up toward unity) in the presence of DNP, and the degree of this suppression varies weakly and nonmonotonically with the quadrupolar angle $\theta$. To understand this behavior better, in Fig.~\ref{fig-pols-omega-e} we show the polarization distributions for five of the minima from Fig.~\ref{fig-feedback-minima} for four different quadrupolar angles. It is clear that for all values of $\theta$, as the electron spin Zeeman frequency due to the external magnetic field, $\omega_{e0}$, is increased, the polarization distributions gravitate toward $m=0$. This is because larger values of the electron Zeeman frequency suppress HF flip-flops, as the violation of energy conservation becomes more pronounced in this case. This is why the $\theta=0$ curve in Fig.~\ref{fig-feedback-minima} monotonically decreases with increasing $\omega_{e0}$. On the other hand, quadrupole-induced nuclear spin flips do not depend on the electron Zeeman frequency, and so these gradually begin to dominate as both $\theta$ and $\omega_{e0}$ increase. This in turn causes the curves in Fig.~\ref{fig-feedback-minima} to become essentially independent of $\omega_{e0}$ as $\theta$ increases. This is another manifestation of how quadrupolar interactions suppress mode-locking effects.

%%%%%%%%%%%%%%%%%%%%%%%%%%%%%%%%%%%%%%%%%%%%%%%%%%
%%%%%%%%%%%%%%%%%%%%%%%%%%%%%%%%%%%%%%%%%%%%%%%%%%
%%%%%%%%%%%%%%%%%%%%%%%%%%%%%%%%%%%%%%%%%%%%%%%%%%
%%%%%%%%%%%%%%%%%%%%%%%%%%%%%%%%%%%%%%%%%%%%%%%%%%

%%%%%%%%%%%%%%%%%%%%%%%%%%%%%%%%%%%%%%%%%%%%%%%%%%
%%%%%%%%%%%%%%%%%%%%%%%%%%%%%%%%%%%%%%%%%%%%%%%%%%
%%%%%%%%%%%%%%%%%%%%%%%%%%%%%%%%%%%%%%%%%%%%%%%%%%
%%%%%%%%%%%%%%%%%%%%%%%%%%%%%%%%%%%%%%%%%%%%%%%%%%

\section{Conclusions}\label{sec-conclusions}

In this work, we developed a general theoretical framework to describe the dynamics of an electron trapped in a self-assembled quantum dot that is driven by a periodic train of optical pulses and coupled to a nuclear spin bath. Using a dynamical, self-consistent, mean-field type approach, we calculated the steady-state dynamic nuclear polarization, as well as its influence on the evolution of the electron spin. Our framework is non-perturbative, applies to nuclei of arbitrary total spin $I$, and includes quadrupolar effects that arise for $I>1/2$.

We showed that the phenomenon of mode-locking, or DNP-induced frequency-focusing, seen in experiments \cite{Greilich2006,Greilich,Carter,Varwig2012,Varwig2013} emerges naturally from our formalism. It can be understood as originating from the structure of the rates for the electron and nuclear spins to flip with one another under the hyperfine interaction. The flip rates vanish when the effective electron precession frequency (including the DNP-driven Overhauser shift) becomes commensurate with the optical pulse train, because in this case the pulses do not interrupt the joint electron-nuclear evolution, and so no polarization is transferred from the electron spin to the nuclei. The vanishing of the flip rates then leads to sharp peaks in the nuclear polarization distribution at magnetization values that satisfy the commensurability condition. Our exact result for the nuclear spin probability distribution in the $I=1/2$ case makes this connection explicit, since the distribution depends inversely on the flip rates.  In addition to mode-locking, we showed that hyperfine flip-flops also give rise to a net nuclear spin polarization that appears to grow linearly with the number of nuclei.

Our formalism includes not only hyperfine-driven phenomena, but also quadrupolar effects that can arise for $I>1/2$. We found that the importance of quadrupolar interactions depends sensitively on the quadrupolar angle $\theta$ between the applied magnetic field and the principal axis of strain in the dot. For $\theta<\pi/8$, hyperfine interactions tend to dominate, leading to clear signatures of mode-locking. However, for $\theta\ge\pi/8$, quadrupole-induced nuclear spin flips begin to dominate, which leads to a suppression of mode-locking and a reduction of the net nuclear polarization. We also showed that quadrupolar effects become more pronounced when the applied magnetic field is increased, because hyperfine flip-flops are suppressed by the increasingly large Zeeman energy mismatch between the electron and nuclei. These effects are clearly visible in the nuclear spin polarization distributions for both $I=1$ and $I=3/2$, and they translate to experimentally detectable signatures that are encoded in the presence or absence of electron spin oscillations in the steady state. Hyperfine flip-flops lead to coherent oscillations in the vicinity of each pulse and halfway between pulses, while quadrupolar interactions act to suppress these oscillations. These signatures offer a potential method to measure the strength of quadrupolar interactions in quantum dots.

The framework we have presented constitutes an efficient, quantitative approach to describing the dynamics of a driven spin coupled to a spin bath. Going forward, it would be interesting to see if some of the simplifying assumptions made here can be relaxed to enhance quantitative accuracy. For example, can we go beyond the box model limit and allow for non-uniform hyperfine couplings, perhaps using a ``wedding cake" model in which the electronic wavefunction envelope is approximated by a piecewise-constant function? Such a generalization would also allow for the inclusion of multiple nuclear species, which is relevant for common semiconductor QD compounds such as InGaAs. It would also be interesting to extend this method beyond the independent nuclei approximation, perhaps using a cluster-based approach in which inter-nuclear interactions are included gradually within clusters of increasing size \cite{Witzel_PRB06,Yao2006}. In terms of applications, our framework could be employed to design driving protocols to achieve desired bath polarization states to either mitigate decoherence or utilize the bath as a quantum memory \cite{Taylor,Gangloff62,Denning}. Finally, we note that the theory we developed is quite general and could be applied to other problems involving a driven system coupled to a quantum bath.

%%%%%%%%%%%%%%%%%%%%%%%%%%%%%%%%%%%%%%%%%%%%%%%%%%
%%%%%%%%%%%%%%%%%%%%%%%%%%%%%%%%%%%%%%%%%%%%%%%%%%
%%%%%%%%%%%%%%%%%%%%%%%%%%%%%%%%%%%%%%%%%%%%%%%%%%
%%%%%%%%%%%%%%%%%%%%%%%%%%%%%%%%%%%%%%%%%%%%%%%%%%

\section*{Acknowledgements}

EB acknowledges support from NSF Grant No. 1847078. SEE acknowledges support from NSF Grant No. 1839056.

%%%%%%%%%%%%%%%%%%%%%%%%%%%%%%%%%%%%%%%%%%%%%%%%%%
%%%%%%%%%%%%%%%%%%%%%%%%%%%%%%%%%%%%%%%%%%%%%%%%%%
%%%%%%%%%%%%%%%%%%%%%%%%%%%%%%%%%%%%%%%%%%%%%%%%%%
%%%%%%%%%%%%%%%%%%%%%%%%%%%%%%%%%%%%%%%%%%%%%%%%%%

\bibliography{biblo}

\end{document}